\documentclass[manuscript,screen]{acmart}
\AtBeginDocument{%
	\providecommand\BibTeX{{%
			\normalfont B\kern-0.5em{\scshape i\kern-0.25em b}\kern-0.8em\TeX}}}

\setcopyright{acmcopyright}
\copyrightyear{2024}
\acmYear{2024}

\acmJournal{TOSEM}

\renewcommand\footnotetextcopyrightpermission[1]{}

\usepackage{algorithmic}
\usepackage{textcomp}
\usepackage{xcolor}
\usepackage[normalem]{ulem}
\usepackage{graphicx}
\usepackage{hyperref}
\usepackage{hyperxmp} 

\usepackage{float}
\usepackage[many]{tcolorbox}
\usepackage{pifont}
\usepackage{wrapfig}
\usepackage{bm}
\usepackage{multirow}
\usepackage{multicol}
\usepackage[skip=1pt]{caption}
\usepackage{color}
\usepackage{colortbl}
\usepackage{enumitem}
\usepackage{threeparttable}
\usepackage{soul}  
\usepackage{setspace}
\usepackage{balance}

\usepackage[ruled,vlined,linesnumbered]{algorithm2e}
\usepackage{array}
\usepackage{calligra}
\usepackage{makecell}
\usepackage{booktabs}

\definecolor{mygray}{gray}{.9}
\definecolor{mypink}{rgb}{.99,.91,.95}
\definecolor{mycyan}{cmyk}{.3,0,0,0}

\AtBeginDocument{%
  \providecommand\BibTeX{{%
    \normalfont B\kern-0.5em{\scshape i\kern-0.25em b}\kern-0.8em\TeX}}}

\newcommand{\find}[1]{
\begin{tcolorbox}[tile,size=fbox,boxsep=1mm,boxrule=0pt,top=0pt,bottom=0pt,sharp corners=northwest,
borderline={0.6mm}{0pt}{black!66!white},colback=black!5!white]
\em #1
\end{tcolorbox}
}

\newboolean{showcomments}
\setboolean{showcomments}{true}
\setboolean{showcomments}{false}
\ifthenelse{\boolean{showcomments}}
 { \newcommand{\mynote}[2]{
      \fbox{\bfseries\sffamily\scriptsize#1}
        {\small$\blacktriangleright$\textsf{\emph{#2}}$\blacktriangleleft$}}}
        { \newcommand{\mynote}[2]{}}

\definecolor{DarkOrange}{rgb}{0.8,0.3,0.0} 
\definecolor{DarkCyan}{rgb}{0.0, 0.55, 0.55}
\definecolor{codegreen}{rgb}{0,0.6,0}
\definecolor{codegray}{rgb}{0.5,0.5,0.5}
\definecolor{codepurple}{rgb}{0.58,0,0.82}
\definecolor{backcolour}{rgb}{0.95,0.95,0.92}

\ifthenelse{\boolean{showcomments}}
{

\newcommand{\sw}[1]{\textcolor{green}{{\it [Cai says: #1]}}}
\newcommand{\yu}[1]{\textcolor{blue}{{\it [Yu: #1]}}}
\newcommand{\swsecond}[1]{\textcolor{red}{#1]}}

\newcommand{\hrmark}[1]{\textcolor{purple}{{[Haoran's comment: #1]}}}
\newcommand{\hrdel}[1]{\textcolor{orange}{{\sout{ #1}}}}
\newcommand{\ques}[1]{\textcolor{blue}{{[Haoran's question: #1]}}}
}
{
\newcommand{\yu}[1]{}

\newcommand{\sw}[1]{}
\newcommand{\swsecond}[1]{}

\newcommand{\hrmark}[1]{}
\newcommand{\hrdel}[1]{}
\newcommand{\ques}[1]{}
}

\def\BibTeX{{\rm B\kern-.05em{\sc i\kern-.025em b}\kern-.08em
    T\kern-.1667em\lower.7ex\hbox{E}\kern-.125emX}}

\begin{document}
\title{A Comprehensive Study of Native Code Bugs in Python Applications}
%
    \author{Haoran Yang}
        \orcid{0000-0001-9298-9757}
	\affiliation{%
		\institution{Washington State University}
		\state{Washington}
		\postcode{99163}
		\city{Pullman}
		\country{USA}
	}
	\email{haoran.yang2@wsu.edu}

 \author{Haipeng Cai}
	\orcid{0000-0002-5224-9970}
	\affiliation{%
		\institution{Washington State University}
		\state{Washington}
		\postcode{99163}
		\city{Pullman}
		\country{USA}
	}
	\email{haipeng.cai@wsu.edu}




\begin{abstract}
The impact of Python applications has been evidenced by their widespread presence in some of the most impactful software domains, such as machine learning frameworks and scientific computing platforms. These applications often integrate native code components written in a lower-level programming language like C. This multilingual construction brings various benefits such as greater performance efficiency and easier interoperability with diverse runtime environments. However, bugs 
in the native code (i.e., \textit{native code bugs}), which are usually stealthy, also constitute a major additional challenge to the quality of the Python applications as a whole. 
Yet despite existing relevant studies, there remains a lack of comprehensive understanding of native code bugs in Python applications. 
In this paper, we aim to mitigate this knowledge gap through the first in-depth study of such bugs, dissecting their common \textit{symptoms}, introducing \textit{locations}, 
\textit{manifestation} characteristics, \textit{root causes}, and \textit{fixes}.
Based on our extensive automated and manual analyses of 216 native code bugs in real-world Python projects on GitHub, we obtained novel findings about and new insights into the 
occurrence mechanisms and resolution strategies of those bugs. 
Among others, we observed that (1) Python native bugs are most commonly introduced at foreign function callsites while symptomized primarily as crashes and abnormal system termination, 
(2) these bugs are often manifested in user-level memory management related operations, while predominantly caused by memory safety and type errors associated with foreign data objects, and (3) those bugs are commonly fixed through correcting memory (e.g., reference count) management 
or incorporating necessary input/boundary checking and validations, often at foreign function callsites or branching control structures. 
We also identified the semantics differences and interactions between the native and host languages as the primary challenge to native code bug detection, followed by the complex ways native code interacts with the foreign language. 

These results further led to actionable suggestions on practical approaches to native code bug prevention, detection, testing/debugging, and fixing. 
For instance, (1) automatic API recommendations focusing on foreign function usage can be of peculiar merits to developers of native Python applications, 
(2) native code behaviors for memory management, foreign object manipulation, and cross-language exception handling should be main testing/debugging targets to exercise for testers of those applications, while 
(3) 
foreign function callsites may be particularly attended when identifying fixes of bugs appearing at those locations, whereas bugs occurring at assignment statements are most likely not to be fixed at those statements.


\end{abstract}

\maketitle



\section{Introduction}\label{sec:intro}
%
Python has stood out as one of the most popular programming languages~\cite{top10language,top20language} thanks to its simplicity and ease of use, among its other merits. 
As a result, Python applications are ubiquitous: for example, almost all of the real-world machine learning (ML) applications are Python applications---mainly because the most dominating ML frameworks/libraries (e.g., TensorFlow~\cite{tensorflow} and PyTorch~\cite{pytorch}) that these applications are built upon are also developed in Python. 
Meanwhile, most of these impactful Python applications and their underlying frameworks/libraries are \textit{multilingual}, i.e., they include integral parts written in other programming languages---notably, \textit{native} languages like C/C++. 
In fact, Python's widespread adoption across various domains has led to an increased reliance on its extensive ecosystem of \textit{third-party} libraries and modules (e.g., SciPy~\cite{SciPy} and NumPy~\cite{NumPy}), which are often backed by native (C/C++) code. For instance, SciPy has 23\% of its code being native code while NumPy has 38\%. PyTorch and TensorFlow have even larger portions implemented in native code: 42\% and 57\%, respectively. 
Even the runtime system of Python itself has a heavy inclusion of native code---although the Python interpreter is written in C, its standard/runtime libraries are written in Python while being supported by native code underneath. 
The Python-C combination is also one of the top language combinations in the world of multi-language software~\cite{li2023multilingual}. 

Generally, the integration of multiple programming languages within a single software project, a practice termed {\textit{multilingual development}}~\cite{yang2024multi}, has become widespread~\cite{jones2010software,delorey2007programming,ray2014large} and continues to experience a rising trend~\cite{tomassetti2014empirical,mayer2015empirical,valverde2015punctuated,li2021understanding}. As a consequence, the \textit{vast majority} of contemporary real-world systems exhibit a {\textit{multilingual}} nature, highlighting the critical importance of assuring their quality. 
However, while facilitating a productive software development process through various means~\cite{abidi2019behind,bissyande2013popularity}, such as leveraging the strengths of individual languages to promote code reuse~\cite{meyerovich2013empirical,vasilescu2013babel}, the practice of multilingual development simultaneously introduces heightened complexity to the resulting software systems. This increased complexity can potentially manifest as a higher propensity for bugs and defects within the resulting multilingual codebase~\cite{abidi2021multi,kochhar2016large,grichi2020impactjni}. 
This is true of multilingual Python applications. 


Indeed, accompanied by its popularity and impact, Python applications with native code in C (i.e., \textit{native Python applications}), have been a prominent case of this phenomenon of multilingual construction leading to additional, often complex code defects in general~\cite{kochhar2016large,grichi2020impactjni}. 
Recent studies of bugs in Python-C applications~\cite{li2023understandinga,li2023understandingb,hu2023empirical} and techniques that are developed to discover such bugs~\cite{wen23usenixsecurity,wen22usenixsecurity,li2023pyrtfuzz,hu2023cross} have revealed the prevalence of interoperability bugs~\cite{chisnall2013challenge} induced by the interactions~\cite{mayer2017taxonomy,wen22fsetool} between Python and C. 
In addition, these prior works further demonstrated that many of the multilingual bugs drill down to bugs that occur in native code (i.e., \textit{native code bugs}). 
%
Thus, 
it is crucial to understand those bugs (e.g., regarding
\textit{where} and \textit{why} they occur as well as \textit{how} to fix them) as 
an essential first step toward the quality assurance of the Python applications holistically. 

However, relevant early studies typically address interoperability issues broadly, such as the misuse of language interfacing APIs~\cite{li2023understandinga,sultana2016understanding,hu2023empirical}, 
faulty cross-language dependencies~\cite{li2023understandingb,grichi2020impactjni},
and violations of interoperation specifications~\cite{bae2019towards,lee2020broadening,kondoh2008finding,li2009finding}. 
Other studies look at the resolution of multilingual bugs~\cite{li2023understandinga} and various multilingual software development issues (e.g., exception handling across languages and choices of language interfacing mechanisms)~\cite{yang2023demystifying,yang2024multi}. 
Studies focusing on native code bugs do exist, yet the extant works fall far short in at least two folds. 
First, they are mostly each only concerned with a single type of (e.g., reference-counting~\cite{li2014finding,ma2023detecting}, exception-checking~\cite{li2011jet}, and interfacing efficiency~\cite{tan2021toward}) bugs, instead of comprehensively examining native code bugs.of various kinds. 
Second, most of these studies on native code bugs are focused on those in the native C code interacting with Java as the host language via Java native interface (JNI)~\cite{kondoh2008finding,li2011jet,grichi2020impactjni,li2009finding,tan2008empirical}. 
Bugs in native code in Python applications are rarely studied and the few extant ones again are exclusively focused on either reference-counting errors~\cite{li2014finding,ma2023detecting} or bugs induced by inefficient interaction between Python code and native libraries~\cite{tan2021toward}. 
As it stands, there is still a lack of \textit{systematic study on the native code bugs in Python applications}.  

In this paper, we set out to fill this critical gap, focusing on real-world {native code} bugs 
in \textit{native Python applications} (i.e., software written in Python and C as primary languages, a.k.a {\em Python-C} software). 
We start by collecting and labeling the first set of 1,039 automatically mined and then manually confirmed native-bug-fixing commits 
from 200 Python-C projects on GitHub.
Then, we conducted a comprehensive study of 216 native code bugs randomly sampled from those commits, systematically examining their \textit{symptoms}, \textit{introducing locations} (in terms of structural/syntactic code contexts), \textit{manifestation characteristics} (concerning the type and semantics of \textit{foreign functions} and the direction/complexity of inter-language information flow), \textit{root causes}, and \textit{fixing strategies}. We also aim to identify key \textit{challenges} to native bug detection, especially those to automated approaches to that task. 


\vspace{2pt}\noindent
\textbf{Major findings.}
Our study reveals that native code bugs are prevalent in Python applications, which are also stealthy and of diverse kinds. We provide novel findings about the common patterns of those bugs, their externalized consequences, underlying causes, and 
mitigating strategies. More specifically, among other observations, we found that 
\begin{itemize}[leftmargin=16pt,topsep=2pt]
\item 
    The native code bugs studied are much more commonly (32.4\%) externalized as abnormal termination of the native Python applications than any other symptoms (e.g., only 7.4\% as error messages yet without the application getting terminated, as the second most common symptom); these abnormal termination cases are demonstrated in forms of application crashes and unexpected exits, resulting primarily (74.3\%) from memory safety issues. 
\item 
    The vast majority (90\%) of these native code bugs appear within a function, with the rest typically falling in a global area (e.g., global variable declaration along with/without initialization) of the native code. The within-function cases are dominated by those occurring at foreign function (i.e., functions invoked in C for accessing Python capabilities) callsites (21.8\%), callsites targeting other functions (14.4\%; i.e., C functions called at cross-language information flow paths associated with the native code bug), and assignments (13.4\%), followed by true/false branches (11.1\%), return sites (10.6\%), and other relatively minor ones. 
\item The studied bugs were most commonly manifested in cross-language semantics related to user-level memory management such as reference counting (41.7\%), foreign object manipulation (31.5\%), and inter-language exception handling (31.5\%). These bugs are by far dominated by the cases (92.1\%) in which the associated cross-language information flow goes from the native (i.e., C) to the foreign (i.e., Python) language, mainly due to the characteristics of these two languages in terms of their type systems and memory management mechanisms. 
\item These native code bugs were usually rooted in memory safety violations (29.17\%), logic flaws (11.57\%), and data type errors (11.11\%), as exhibited in missing boundary check against calls to foreign functions and inconsistent data types returned by foreign function calls. 
The dominating root causes, memory safety violations, are primarily attributed to reference count misuse (38.1\%) and null pointer dereference (31.7\%). 
%
%
%
%
%
\item The dominating technical challenge to automated Python native code bug detection (52.8\%) lies in the semantics differences and interactions between the native and host languages, followed by the complexity in foreign language interfacing (25.9\%) as the second most common challenge. 
\item For fixing these native code bugs, code changes are almost always (99.1\%) made within a function versus (1\%) in the global area of the native code, mostly at a foreign function callsite (36.0\%), a whole branching (e.g., {\tt If}) structure (16.8\%), and conditional expression (12.6\%).
Fixing strategies for native code bugs tend to be diverse. 
The relatively common 
fixing strategies 
are 
correcting memory management or reference counting (35.2\%), incorporating boundary
checks or input validations (20.8\%), and applying appropriate error or exception handling (12.0\%).
\end{itemize}

\noindent
Furthermore, we propose guidelines and best practices for mitigating native code bugs in Python applications, through prevention, testing/debugging, and fixing. Our findings raise awareness of native code bugs and their impact, while highlighting the need for better tooling support and techniques to manage the code defects associated with native code integration in Python projects. 

\vspace{2pt}\noindent
\textbf{Contributions.}
In summary, our study makes the following main contributions: 
\begin{itemize}[leftmargin=16pt,topsep=2pt]
\item To the best of our knowledge, this is the {\em first comprehensive study} of real-world native code bugs that addresses \textit{where} they are introduced and \textit{why} they occur, as well as \textit{how} they are manifested and fixed. 
\item Our {\em novel empirical findings} can help researchers and developers gain a systematic and in-depth understanding of these bugs hence guide their countermeasures.  
\item Based on these findings, we offer suggestions and insights on how to avoid, detect, test/debug, and repair native code bugs in Python applications, which can inform design of tooling support of respective capabilities. 
\item We also contribute the first high-quality (accurate and diverse) {\em dataset} of native code bugs which can be immediately reused by future researchers to conduct further relevant studies and develop Python application assurance tools. 
\end{itemize}






\vspace{2pt}\noindent
\textbf{Artifact.}
All of the source code and datasets used in and produced from our study have been made publicly available at
\href{https://figshare.com/s/995c9d9bbaaab500b6cf}{\underline{https://figshare.com/s/995c9d9bbaaab500b6cf}}.

\section{Background}\label{sec:background}
This section defines key concepts and terminologies in
the rest of this paper that are necessary for understanding it.

\vspace{2pt}\noindent
\textbf{Multi-language software}. 
A software system developed in multiple computer languages consists of two or more code units each written in one single language, referred to as a \textit{language unit}~\cite{haoran22fsenier}. 
The languages utilized can be of various classes, such as programming languages (e.g., traditional ones like C and scripting languages like JavaScript), modeling languages (e.g., markup languages like HTML and query languages like SQL), and descriptive languages (e.g., UML and BPMN). 
However, such a system is considered \textit{multi-language software} only if the language units interact with each other~\cite{li2021understanding}. 
Without loss of generality, we will use multi-language software and \textit{multilingual software} exchangeably. 

\vspace{2pt}\noindent
\textbf{Language interfacing}.
Accompanying the diversity of languages is the variety of ways in which 
the different language units of multilingual software interact/interface among them; and the way is referred to as \textit{language interfacing mechanism}~\cite{wen22fsetool}. 
Among the various kinds of such interfacing mechanisms, 
\textit{Foreign Function Interface (FFI)} is one provided by a programming language that allow code written in that language to call functions or procedures written in another language~\cite{wen22fse,lee2020broadening,grichi2020impactjni}. 
FFI is the primary mechanism by which two \textit{programming} languages interact, 
including between Python and C~\cite{wen22usenixsecurity,wen23usenixsecurity,wen22fse,youn2023declarative}. 
More generally, it is typically used when developers want to call a function in a library written in a different language, or when they need to interface with low-level system APIs that are exposed through another language like C. 
For Python-C projects, for example, commonly used FFI instances are 
 Cython~\cite{cythonsite}, ctypes~\cite{ctypesweb}, cffi~\cite{cffiweb}, and 
 SWIG~\cite{swigweb}. 

\vspace{2pt}\noindent
\textbf{Cross-language function/API}.
When using an FFI for language interfacing, 
the interoperation between the two languages is realized by the FFI via 
a set of functions each written in one language $L1$ but either \textbf{(a)} to be called in another language $L2$ or to \textbf{(b)} serve the purpose of accessing (e.g., the capability offered by or data originated in) $L2$ from within a $L1$ unit.  
Each of these functions is referred to as a \textit{cross-language function}. 
Especially, if a cross-language function is exposed/externalized for public use by the language other than the one it is written in, we also refer to it as a \textit{cross-language API}. 
Non-API cross-language functions 
are not uncommon---they often exist to serve as helper functions 
to immediately support the functionalities of related cross-language APIs. 

Given an FFI, there are two main classes of cross-language APIs: 
native functions and foreign functions~\cite{yang2023demystifying}. 
In the situation (a) above, the API is typically written by the application developer in $L1$ but called in $L2$; this API is considered a \textit{native function} and accordingly $L1$ is considered a \textit{native language}. 
In the situation (b) above, the API is usually written by the language ($L2$)'s developer in $L1$ and also to be called in $L1$; this API is considered a \textit{foreign function} and $L2$ is considered a \textit{foreign language} or \textit{host language}. 
In a Python-C project, C is the native language and Python is the foreign/host language. 


\vspace{2pt}\noindent
\textbf{Native code bug.}
In a multilingual software project, the code unit written in the native language is referred to as the \ul{native code}, while 
the code unit written in the foreign/host language is referred to as the foreign/host code. 
Moreover, in such projects, the developers may encounter various kinds of bugs in terms of language scoping. Some of the bugs are constrained in the foreign/host code, while others occur in the native code. 
Bugs that appear in the native code are referred to as \ul{native code bugs}, or simply \ul{native bugs}.
In particular, in a Python-C project, the C code is the native code; hence, the native code bugs are those that appear/manifest in the C code. 
\section{Methodology}\label{sec:methodology}
\begin{figure}[tp]
	\vspace{-0pt}
	\includegraphics[width=0.85\linewidth]{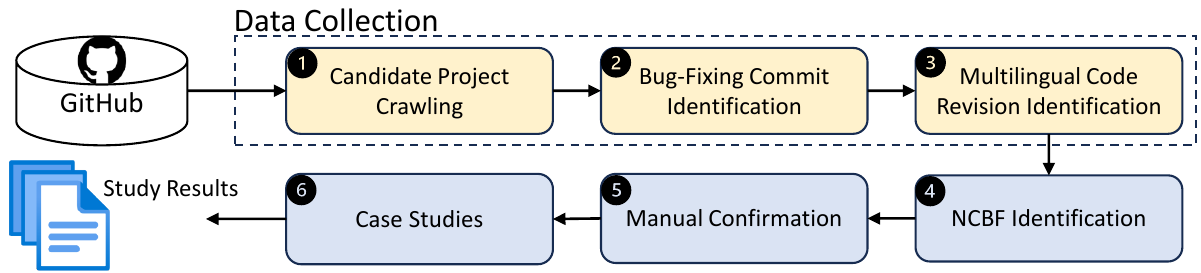}
	\caption{Overview of our study methodology (process).}
	\label{fig:studyprocess}
	\vspace{-12pt}
\end{figure}
In this section, we present our study process as sketched up in Figure~\ref{fig:studyprocess}. 
Since we aim to understand native code bugs, we 
need to first collect datasets on such bugs. 
Some seemingly relevant datasets exist, such as Defects4J~\cite{just2014defects4j} for Java bugs, Codeflaws~\cite{tan2017codeflaws} for C bugs, and BugsInPy~\cite{widyasari2020bugsinpy} for Python bugs. However, these prior datasets are 
all of bugs in single-language software---hence, the bugs are not native code bugs.

Given its open, large, and diverse information on 
real-world software projects, we chose GitHub as the source of our data collection, retrieving native code bugs from 
the bug-fixing commits available in its (multilingual) software repositories. 
Yet not all of the GitHub projects are for multilingual software and not of all the commits in the multilingual projects are for fixing native code bugs, making the data collection and labeling a challenging task.

Specifically, as Figure~\ref{fig:studyprocess} shows, we start with 
(1) \ul{candidate project crawling} on \textit{GitHub} to obtain the \textit{list of (potential) multilingual repositories} 
and all of the \textit{associated commits} with them. 
Then, (2) from this initial list and these commits, 
we collect 
code commits that aim to fix a bug through \ul{bug-fixing commit identification}.
This is followed by (3) \ul{multilingual code revision identification} to 
ensure that the revision corresponding to a bug-fixing commit 
is multilingual software. 
We proceed with identifying bug-fixing commits that actually aim to fix \textit{native code bugs}, noted as native-code-bug-fixing \ul{(NCBF) identification}.  
Due to the imprecision of previous steps, we then finalize on NCBF commits through \ul{manual confirmation}, aiming to remove false positives. 
From these confirmed 
NCBF 
commits, we obtain the native code bugs and only keep the associated NCBF commits. 
Finally, guided by predefined \textit{research questions}, 
we analyze the confirmed \textit{native code bugs} through \ul{case studies} 
to obtain the \textit{study results}.

\subsection{Data Collection}\label{ss:repomining}
This section describes our data collection process, used for gathering multilingual project repositories that contain potential NCBF commits hence native code bugs. 
It involves three main steps: candidate project crawling, bug-fixing commit identification, and multilingual code revision identification.
\subsubsection{Candidate Project Crawling}
Multi-language software projects are quite diverse in terms of their used languages~\cite{ray2014large,wen22fse,li2021understanding}. 
Thus, understanding native code bugs both comprehensively and in depth
while covering different language combinations would be difficult 
in one study/paper. 
We took Python-C as a starting point for three reasons. 
First, Python and C are both consistently top-10 popular languages, 
and 
have large, active communities that provide support and resources for developers. 
Second, C is a low-level language known for its 
efficiency merits while Python is a high-level language known for its ease 
of use/programming. 
Third, Python and C, when combined, have a large software ecosystem that backs up various computing tasks. 
In particular, leading AI/ML frameworks and the huge corpus of applications based on them, are Python-C software~\cite{grichi2020impact}. 
Thus, Python-C is one of the most impactful language combinations~\cite{wen22fse,wen22usenixsecurity,wen23usenixsecurity} in the multi-language software world.  


For every potential project, we utilize the GitHub API~\cite{github_api} to gather data on the languages employed and the code size for each language component. Utilizing this data, we identify Python-C projects based on three criteria, which are outlined and rationalized as follows:
\begin{itemize}[leftmargin=16pt,topsep=2pt]
\item \textbf{(1)} 70\% or more of the total code base of the project, measured in Source Lines of Code (SLOC), is composed of Python and C together---such projects can be appropriately and reasonably considered Python-C software~\cite{wen22usenixsecurity,wen23usenixsecurity}.
\item 
\textbf{(2)} Both C and Python each accounts for 10\% or more of the entire project in terms of code size (SLOC)---such projects are more likely to have substantial inter-language behaviors hence native code bugs than otherwise per our preliminary experiments. 
\item 
\textbf{(3)} 
The project's repository is marked by a certain number of stars, $n$, a metric that previous research~\cite{bonifro2021content,di2020multinomial,romano2021g,ray2014large,papamichail2016user,wen22fse,li2021understanding} has frequently utilized to measure a project's popularity. Projects with a higher $n$ are typically more noticed, meaning any bugs they contain could have broader impacts. Therefore, a higher $n$ is desirable. However, an increase in $n$ limits the number of projects available for collection. For our research, having access to a broader range and number of projects would render the results of our study more comprehensive. With this trade-off in mind and per our preliminary repository mining outcome, we ultimately decided on $n=10$ as an optimal compromise between these two competing factors.
\end{itemize}

Following these criteria, we obtained a list of 2,613 repositories. 
The next step is to mine commits in each repository. 
We considered all the available commits in these repositories, amounting to a total of 652,294 commits. 



\subsubsection{Bug-Fixing Commit Identification}\label{sss:bfci}
Intuitively, an NCBF commit is required to be a commit that fixes a bug. However, the commits we gathered at the conclusion of our repository mining process, as detailed in Section $S$\ref{ss:repomining}, can include a variety of types, such as commits for new feature additions or even changes that do not involve code. Apparently, manually filtering through the vast number of commits we collected is not practical.

When making a commit, developers might link it to the corresponding issue (id) that the commit resolves. Furthermore, each issue typically comes with a label or tag that categorizes it according to its nature. Given that not every issue pertains to code bugs, we employ a set of keywords related to bugs to screen the issue tags. These keywords include 'error', 'bug', 'defect', 'patch', 'mistake', 'fault', 'failure', 'fix', 'issue', 'incorrect', and 'flaw'. A commit is regarded as a bug-fixing commit if it is connected to an issue tagged with any of these keywords. However, developers do not always link a bug-fixing commit to an issue, potentially leading to the miss of some actual bug-fixing commits in our collection. Nevertheless, this method offers high precision (e.g., more than those based on matching commit messages against such keywords~\cite{ray2016naturalness}). In our study, we choose to emphasize precision over recall, as our objective is to analyze native code bugs accurately identified as such. By the conclusion of this process, we had identified 32,672 bug-fixing commits.

\subsubsection{Multilingual Code Revision Identification}\label{sss:mcri}
A commit to the repository of a project is an NCBF commit only 
if the project is multilingual. 
Yet the mere use of two languages in a project as checked during our repository mining does not necessarily indicate that the project is multilingual---in some cases, a project simply implements the same functionalities with different languages to accommodate the needs of various user groups; for our study, the two languages must also have interactions between them---otherwise, the C code in the project may not even be native code, hence ruling out the candidacy of the project for our native code bug study. 
Moreover, checking such interactions at project level may not suffice because the language use and multilingual nature of a project may change across revisions~\cite{li2021understanding,wen22fse,li2023multilingual}. 
For instance, we observed that some projects are multilingual (e.g., using ctypes for language interfacing) at one point (with respect to a revision/commit), but not at a later time (i.e., another revision/commit). 

Thus, we check for each commit if the corresponding \textit{code revision} is multilingual. To that end, we developed an automaton-based utility
to detect whether a project revision has language interactions via common Python-C FFIs 
(i.e., {\tt ctypes}, {\tt Cython}, {\tt cffi}, {\tt C extension}, and {\tt SWIG}) 
and, if any, classify the interaction types.  
Our tool is based on PolyFax~\cite{wen22fsetool}, which supports only the first two types of FFI for Python-C projects. We extended it to support the other three types in order to comprehensively identify/classify any Python-C interaction in the given repository revision. 
After this step, we have 6,981 commits. 

\subsection{NCBF Identification}\label{sss:mbfci}
One prerequisite for the existence of NCBF in a multilingual project revision is that there are associated commits that are aimed to fix a bug in the revision. 
Yet given a commit that is associated with fixing a bug in a multilingual project (revision), the commit may still not necessarily aim to fix a \textit{native code bug} (e.g., it may target a single-language bug entirely contained in Python, or a cross-language bug rooted in Python). 

To ensure the commit is actually associated with a \textit{native} bug, 
the changed code should be part of the 
native code unit of the revision (which has been confirmed as multilingual). 
Also, the changed code must also \textit{potentially}\footnote{Some of the native code may have not yet been, but can potentially be, invoked by the Python code.} impact cross-language behaviors 
of the revision---such behaviors can be modeled by cross-language information flows/dependencies. 
Moreover, in an FFI-based Python-C revision, language interactions are realized through native and
foreign functions. 
Thus, a bug-fixing commit in a multilingual project (revision) is an NCBF commit 
if at least one of the changed code lines (1) has a (transitive) dependence (backward or forward) relationship 
with at least one of the foreign function callsites in the native code unit, or (2) enclosed either directly in a native function definition or in the definition of a function that is reachable from a native function definition via those dependencies (transitively).

However, a static information flow/dependence analysis tool widely applicable to real-world Python-C software is not available~\cite{youn2023declarative}. 
The closest available one is Joern~\cite{yamaguchi2014modeling}, which supports intraprocedural control flow analysis for both Python and C. 
Based on these capabilities, we extended Joern to build the whole-program interprocedural control flow graph (ICFG) for a given project (revision). 
Then, according to the FFI type identified in its previous step ($\S$\ref{sss:mcri}), 
the module retrieves the foreign functions for that type, including those crawled from the FFI's official documentation (e.g.,~\cite{ctypesweb} for ctypes) and native functions extracted from respective interfacing (e.g., C header) files. 
Finally, we consider a given commit \textit{may} be an NCBF commit 
if at least one of the changed code lines meets (1) or (2) above. 
The rationale under ``may" here is that control flow reachability between two program entities $a$ and $b$ is a necessary (albeit not sufficient) condition of any information flow/dependence relationship between $a$ and $b$. 
In this way, the ICFG-based analysis is a reasonably safe approximation in identifying a commit as an NCBF or not.
After this step, we have 1,751 (\textit{likely} NCBF) commits. 

\subsection{Manual Confirmation}\label{ss:manconfirm}
Through the data collection and NCBF identification, the resulting commits are still likely to be false positives as NCBF commits, due to the conservative nature of the control flow approximation discussed above. 
Thus, we proceed with a manual confirmation process, during which 
we manually check if each of those resulting commits is a true NCBF commit. 
Specifically, given such a commit, we follow each control flow path between any of the commit change locations and one of its reachable foreign function callsites, as well as ICFG paths between those locations and code entities in any native function definition or the definition of any function that is reachable from a native function definition. 
Essentially, this is to identify if the changes impact any of such callsites or entities through data/control dependencies along the respective control-flow paths, if any. 
If the impact exists, we confirm the commit as an NCBF commit and include it in our final NCBF commit set, and dismiss it from that set otherwise. 
Three authors made initial decisions for each commit independently, followed by meetings/discussions to resolve disagreements and reach a consensus. 
After this manual analysis, we have 1,039 confirmed NCBF commits.

\subsection{Research Questions}\label{ss:rqs}
The overarching goal of our study is to offer a comprehensive understanding of real-world native code bugs with respect to 
their \textbf{5 aspects}: \textit{symptom}, \textit{location}, \textit{manifestation}, \textit{root cause}, and \textit{fixing strategy}. 
Accordingly, we formulated the following five research questions, as outlined and justified below: 

\textbf{\textit{RQ1: What are the symptoms of native code bugs?}} 
This RQ aims to understand how these bugs are commonly 
externalized (as observed by developers or users). 
The understanding not only serves an initial step of high-level detection of native code bugs in multilingual projects, but also informs about their severity and consequences. 

\textbf{\textit{RQ2: Where are native code bugs introduced?}} 
This RQ aims to discover where these bugs are introduced in terms of code structural/syntactic contexts. 
The result can shed light on the context in which native code bugs may happen hence helping developers understand where to avoid introducing such bugs.

\textbf{\textit{RQ3: What are the manifestation characteristics of native code bugs?}} 
This RQ aims to elucidate the code-level attributes regarding how these bugs are manifested, concerning 
functional semantics (e.g., memory access versus I/O) of the foreign functions involved.  
In a multilingual program, native code usually takes and then processes data passed from the foreign language (i.e., foreign data objects). 
In a Python-C program in particular, the native code typically relies on calling relevant foreign functions to parse the foreign data objects (e.g., a tuple in Python). 
Similarly, other native code behaviors (e.g., managing memory related to foreign data) also involve respective foreign functions (e.g., administering reference counting). 
Thus, the semantics of these foreign functions helps capture the high-level behaviors of the native code, illuminating the manifestation mechanisms of respective native code bugs. 
The result from answering this RQ can help design effective testing techniques to expose native code bugs. 

\textbf{\textit{RQ4: What are the root causes of native code bugs?}} 
This RQ aims to dissect the underlying causes of these bugs. 
The result provides a basic yet essential understanding of native code bugs
and immediate guidance for preventing, detecting/locating, 
and resolving them in native code of multilingual projects. 

\textbf{\textit{RQ5: How are native code bugs fixed?}} 
This RQ aims to reveal how developers have fixed these bugs. 
The result helps inform the development of better native code debugging/repair tools and improve the patching process. 

Finally, based on our understanding about native code bugs in Python applications as formed through characterizing them via the 5 aspects, we 
further identified the key challenges to detecting such bugs. 

\textbf{\textit{RQ6: What are the challenges to native code bug detection?}}
This RQ aims to identify the difficulties and obstacles in detecting native code bugs. The result provides insights into the current limitations of bug detection techniques and highlights the areas that require further research to address native code bugs effectively.

\subsection{Case Studies}\label{ss:studyprocedure}
To answer our RQs, we manually examine the native code bugs associated with the confirmed NCBF commits via case studies. 
The answers would involve terms/phrases that we had no full prior knowledge about. Thus, we adopted an open coding approach for each RQ, first 
deriving a codebook and then applying it to characterize each bug in each of the 6 aspects---both by three of the authors following 
a common inter-agreement and consensus procedure. 

\textit{\textbf{Random sampling for codebook.}}
To 
create 
the codebook, we first randomly sampled 281 out of the 1,039 commits for analysis. This sample size is statistically significant at 95\% confidence level (CL) and 5\% margin of error (ME). 

\textit{\textbf{Derive codebooks.}}
We need at least one codebook for each RQ. 
For instance, for RQ4, we want a root cause codebook. 
To that end, each of the three authors independently curated an initial catalog of root cause categories based on the 281 commits. 
Following that, they resolved any discrepancies until they reached a consensus on a finalized codebook. Specifically, each participant undertook the steps of (1) carefully examining the commits, (2) determining if each commit fits in any of the pre-established categories, and (3) creating a new category wherever needed.

In the process of formulating a new category, 
the initial step involved defining a label that accurately described the root cause highlighted by the commit. Subsequently, we developed comprehensive descriptions for this newly established category and pinpointed the root cause that would be suitably classified within it. This commit was then added to the codebook, acting as a representative example for the new category.
The codebooks for all of the other aspects of the native code bugs (e.g., symptom) were generated using a similar procedure.

\textit{\textbf{Random sampling for study.}}
Considering the significant amount of time typically required to manually review even a single commit, we randomly sampled 216 commits (in addition to, i.e., not overlapping with, the 281 utilized for developing the codebook) for our final analysis of bugs.
This sample size is statistically significant at 90\% CL and 5\% ME. 

\textit{\textbf{Coding process.}} 
We applied the corresponding codebooks for each RQ to analyze the 216 bugs. To guarantee the reliability of this method, we used a negotiated agreement strategy, commonly chosen to adopt the generation of new insights~\cite{morrissey1974sources}, which is necessary for our study.
\section{Result and Findings}
In this section, we present and discuss main results and findings, as answers to our research questions. 

\subsection{RQ1: Symptoms of Native Code Bugs}\label{RQ1}
In our study, we found five main types of symptoms (\textbf{Sym}) of native code bugs in Python applications:

{\ul{Sym1: Crash/abort (70).}} 
The program stops/exits unexpectedly. 
We also found that the vast majority of these crashes are related to memory safety issues (52), followed by boundary conditional errors (3) and then input/output handling issues (3). 
While this symptom may not be specific to native code bugs, the fact that 
the largest portion of our studied native code bugs exhibiting this symptom brings to attention that native code bugs do cause fatal consequences. 

{\ul{Sym2: Error messages (16).}} 
The program provides clear alerts when it comes across a scenario it cannot manage or whenever an operation does not succeed.
The error messages associated with this symptom involved type errors, import errors, value errors, etc., which are further connected to underlying issues such as data type inconsistencies (6), API/function misuse (2), etc. 

{\ul{Sym3: Data corruption (11).}} 
The program makes unintended changes to the original data. 
The related issues mainly include data type errors (5), logic errors (3), and API/function misuse (1). 

{\ul{Sym4: Performance bottleneck (7).}} 
The overall system's performance or capability is constrained by one or more of its components.
We further found the related issues mainly concern 
resource management errors (4) and API/function misuse (2), which together actually account for almost all of the bugs that demonstrated this symptom.  

{\ul{Sym5: Deadlock/hang (4).}} 
In the program, two or more operations are caught in an endless wait for each other to finish, or a process becomes unresponsive, waiting for a specific condition to be satisfied.
The associated issues are mainly concerned with improper exception handling (2) and 
API/function misuse (1). 

In other cases, we could not determine the symptoms because the issue description/discussion does not provide the necessary information, nor does the commit message or code comment. 
Moreover, we could not find necessary resources for actually running the respective Python application to 
reproduce the bugs (hence observing the symptoms by ourselves). 
There is a clear need for improvement in the reporting/documentation quality (at least via GitHub issues) of native code bugs. 

In all, symptoms of native code bugs do not appear to be unique or specific to Python applications with native code---bugs in single-language (C) 
code often exhibit similar symptoms. 
However, as we revealed later, the associated issues and underlying root causes are often 
indeed specific to native code (e.g., the type errors with foreign data objects and misuse of foreign functions). 

\find{
Symptoms of the studied native code bugs are dominated by \textit{crash/abort} (32.4\%), followed by \textit{error messages} (7.4\%), \textit{data corruption} (5.1\%), and other even minor cases such as \textit{performance bottleneck} and \textit{deadlock/hang}. These symptoms are commonly associated primarily with  memory safety issues, although also with data type errors and API/function misuse. 
}
\subsection{RQ2: Bug Introducing Locations}\label{RQ2}
At a high level, 
the studied native code bugs were introduced either within a function or outside any function. 

\subsubsection{Within Function} 
For the bugs introduced within a function $f$, we found 9 major types of locations (\textbf{Loc}) as the structural/syntactic contexts they occurred in. 

\underline{{Loc1:Foreign function callsite (47).}} 
The bug was introduced directly at a foreign function callsite within the function $f$.
This is the most common type of introducing location of native code bugs. 

\underline{{ Loc2:(Non-foreign) function callsite (31).}} 
The bug occurred at a callsite of a function within $f$, yet the called function itself is not a foreign function or any other cross-language function. However, this function call is part of the cross-language information flow that led to the bug.
The callee is misused or itself faulty. 

\underline{{Loc3:Assignment (29).}}
The bug occurred during an assignment statement where a variable was set to a particular value.
In these cases, the variables often refer to foreign data, immediately or indirectly. 

\underline{{Loc4:True/False branch (24).}}
The bug was introduced within a conditional code block within $f$, occurring either in the true branch (within the {\tt if} body) or the false branch (within the {\tt else} body).

\underline{{Loc5:Return site (23).}}
The bug occurred at a return site within the function $f$. Not all return sites necessarily include a return value being taken in the caller.

\underline{{Loc6:Following a foreign function call (14).}}
The bug was introduced at a statement of $f$ that immediately follows a call to a foreign function. 
Typically, the statement depends on that call. 

\underline{{Loc7:Conditional expression (13).}}
In these cases, the bug appeared in a conditional expression, such as a predicate, located at a conditional statement within the function $f$.
This expression often directly impacts the cross-language control flow underlying the bug. 

\underline{{Loc8:Loop (9).}}
The bug occurred inside an iterative control-flow structure, e.g., within the body of a loop, where there is no native bug falling in a previous type of  bug locations as described above. 
%

\underline{{Other (4).}}
In a small number of cases, the bug was introduced in other, miscellaneous types of locations, such as object instantiation and {\tt struct} declaration. 

\subsubsection{Outside Function}
From the 22 bugs identified as occurring outside of any functions, the majority (19 cases) were found in the global area (e.g., declaring and/or initializing a global variable). 
The native code bugs also arose in function or variable declarations and preprocessor directives, in 3 cases in total. 


\begin{table}[ht]
\centering
\caption{Distribution of studied native code bugs over various types of introducing locations}
\label{fig:introducing}
\begin{tabular}{llr}
\toprule
 & \textbf{Introducing Location} & \textbf{Count (Percentage)} \\
\midrule
\multirow{8}{*}{Within function} & Foreign function callsite & 47 (21.8\%) \\
 & Non-foreign function callsite & 31 (14.4\%) \\
 & Assignment & 29 (13.4\%) \\
 & True/false branch & 24 (11.1\%) \\
 & Return site & 23 (10.6\%) \\
 & Following a foreign function call & 14 (6.5\%) \\
 & Conditional expression & 13 (6.0\%) \\
 & Loop & 9 (4.2\%) \\
 & Other & 4 (1.9\%) \\
\midrule
Outside function & & 22 (10.2\%) \\
\bottomrule
\end{tabular}
\end{table}

As shown in Table~\ref{fig:introducing}, 
out of the 216 cases analyzed, 194 bugs (nearly 90\%) occurred within a function, indicating these bugs are generally tied to functional logic (rather than global areas, for instance). 
Among the within-function bugs, callsites are a dominating location where these bugs occurred. 

\find{
Native code bugs can be introduced in a variety of code locations/structures, but they 
most commonly occurred within a function (90\% of the cases), especially at a foreign function callsite (21.8\% of within-function cases), callsite of 
other functions (14.4\%), and assignment (13.4\%). 
The bugs outside functions are often located in a global area of the native code, where a global variable is declared with/without initialization. 
}
\subsection{RQ3: Manifestation Characteristics}\label{RQ3}
To understand how native code bugs are manifested, we mainly examined their manifestation characteristics in terms of 
the semantics of foreign functions involved. 
%
%
%
Per its nature and definition, as we justify this RQ on manifestation characteristics (\textbf{Mac}) earlier, a native code bug involves calls to responsible foreign functions. 
The semantics of these functions immediately inform the semantic nature of the bug, which mainly fell in 6 categories: 

\underline{{Mac1:Reference counting (88).}}
Reference counting is a fundamental aspect of Python's memory management, helping in the effective allocation and release of memory. Nonetheless, this result suggests that it is a primary type of code semantics via which the studied native code bugs were manifested. Such bugs typically emerge due to incorrect increments or decrements of reference counts, resulting in memory leaks or premature disposal of objects.

\underline{{Mac2:Object manipulation (68).}} 
Another major type of native code bugs related to cross-language semantics involves the manipulation of foreign (Python) objects, such as their creation, access, and management. These issues encompass problems like erroneous object creation, misuse of object methods, or mishandling of object lifecycle. 
This result points to the challenges associated with managing Python objects in the native C code. 

\underline{{Mac3:Exception handling (68).}}
Python's APIs for interfacing with foreign functions provide mechanisms for signaling errors and managing exceptions. Among the bugs identified, 68 were linked to issues with cross-language semantic handling of exceptions. These issues included failing to capture exceptions thrown in Python or improperly relaying exceptions to the exception-throwing Python unit. This number highlights the considerable difficulty in handling erroneous program states between Python and C. 

\underline{{Mac4:Data/format parsing (14).}}
This refers to the process of parsing a data structure based on rules or a specific grammar according to a format string. In our study, bugs in native code often arose from semantic errors induced by inconsistencies or mistakes in how foreign functions processed data from Python or converted Python data into a format that is comprehensible to and compatible with C.

\underline{{Mac5:Protocol processing (5).}}
A few bugs in our study were manifested when the responsible foreign function 
implements or uses network protocols, communication protocols, or protocol-based 
foreign functions. 
This small percentage (2.5\%) suggests the relative rarity of such situations. 

\underline{{Other (10).}}
In these 10 cases, the bug-manifesting foreign functions have 
miscellaneous kinds of semantics. For instance, they include foreign functions that target operating-system-level utilities or 
C runtime 
APIs for importing external modules (e.g., third-party libraries) or setting handlers for asynchronous events.

\find{
The native code bugs studied were mostly manifested in code semantics on cross-language memory management (41.7\%, reference counting in particular), foreign object manipulation (31.5\%), and inter-language error handling (31.5\%), pointing to the prevalence of issues/challenges associated with these native code behaviors. 
%
%
}
\subsection{RQ4: Root Causes of Native Code Bugs}\label{RQ4}
%
To dissect why native code bugs happen, we conduct root cause analysis of the studied bugs by tracing the underlying data/control flow facts of each bug. 
As a result, we identified 13 root causes of native code bugs in Python applications as elaborated as shown in Table~\ref{fig:root-overview}. 
This analysis of root cause types shows that the most prevalent issues are related to memory safety, logic errors, data type errors, and boundary conditional errors. The dominance of memory safety issues, accounting for nearly 30\% of all bugs, emphasizes the importance of correctly managing memory across Python and C. 
Logic errors, representing about 11.57\% of the total bugs, are highlighted when integrating code written in these two languages where complicated data/control flows are error-prone. Data type bugs, comprising 10.65\% of the total, are also common, pointing to the challenge of correctly understanding and handling data types when data is passed between Python and C. Boundary conditional errors, representing about 9.26\% of the total bugs, pertain to erroneous handling of edge cases in native code behaviors. The "other" category in our root cause analysis encapsulates some issues that are difficult to classify furthermore. These are related to general concerns such as code generation, compilation errors, and installation processes.


\begin{table}[ht]
\centering
\caption{Distribution of root causes of native code bugs}
\label{fig:root-overview}
\begin{tabular}{lrc}
\toprule
\textbf{Root Cause} & \textbf{Count} & \textbf{Percentage} \\
\midrule
Memory Safety & 63 & 29.17\% \\
Logic Error & 25 & 11.57\% \\
Data Type Error & 24 & 11.11\% \\
Boundary Conditional Error & 20 & 9.26\% \\
API/Function Misuse & 14 & 6.48\% \\
Improper Exception Handling & 13 & 6.02\% \\
Initialization\&Variable Misuse Errors & 12 & 5.56\% \\
Mathematical Errors & 6 & 2.78\% \\
Input and Output Handling & 6 & 2.78\% \\
Compatibility & 6 & 2.78\% \\
Resource Management Errors & 5 & 2.31\% \\
Access and Permission Errors & 4 & 1.85\% \\
Other & 18 & 8.33\% \\
\bottomrule
\end{tabular}
\end{table}


Next, we elaborate these root causes, as derived from our coding procedure. 
For each root cause (\textbf{RC}), we start with the definition, followed by an in-depth analysis of it and its implications. 
Furthermore, we illustrated the root cause through representative bug cases. 
We focus on the top-4 root causes, which together account for 61.1\% of all of the native code bugs we studied. 

\begin{figure}[tp]
	\vspace{-0pt}
	\includegraphics[width=0.55\linewidth]{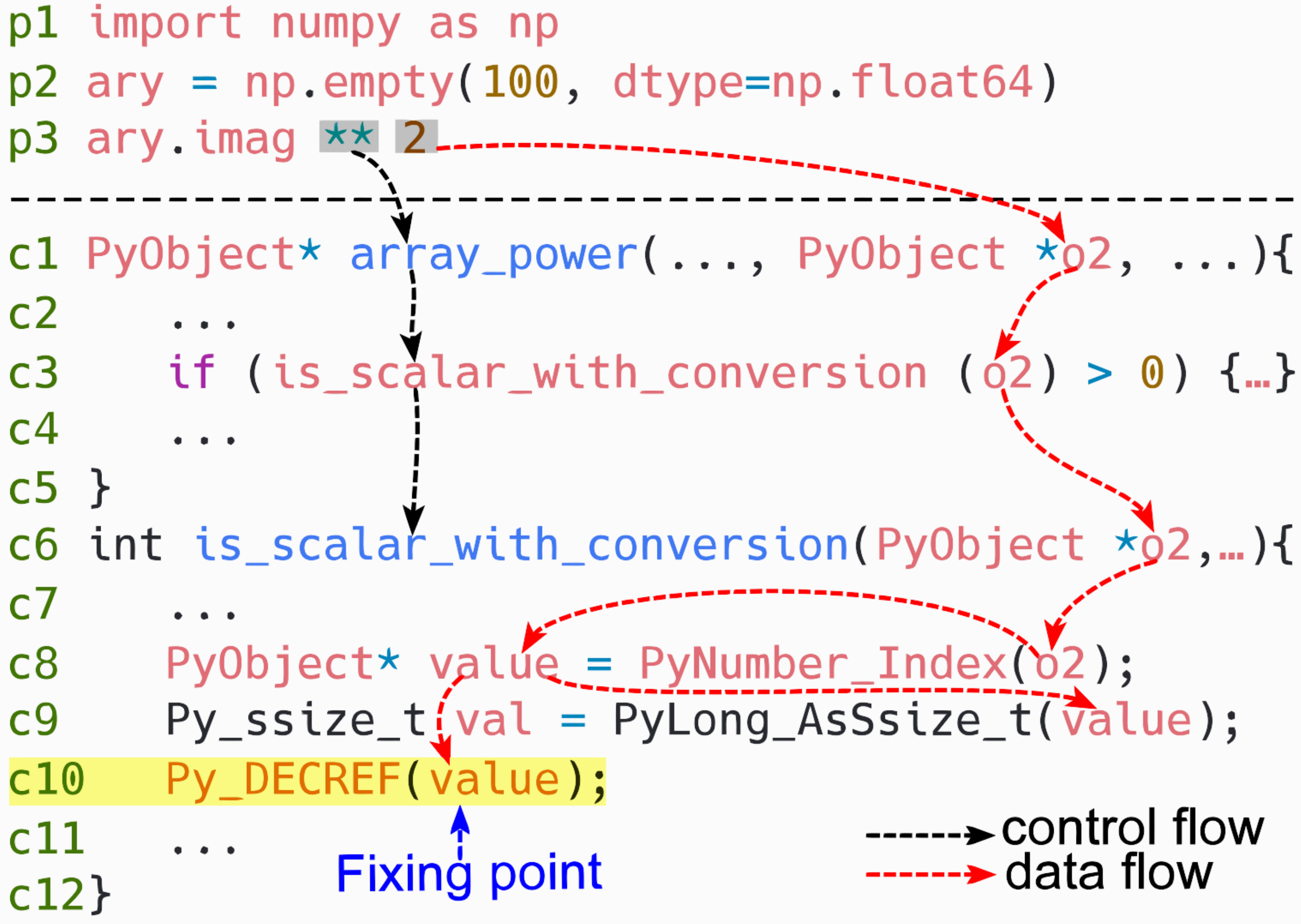}
	\caption{A native bug case rooted in a misused reference count~\cite{casestudy_reference}.}
	\label{fig:root-memory}
	\vspace{-12pt}
\end{figure}

\underline{{RC1: Memory safety (63).}} Memory safety is a critical characteristic of programming that prevents unintentional behaviors due to incorrect memory management. If a program is memory-safe, it efficiently averts bugs and security vulnerabilities where it erroneously accesses memory, allowing it to read from or write to unauthorized memory locations. 
This prevalent root cause may not be surprising, given the capabilities of the C language in accessing and manipulating memory. 

Given the substantial quantity of bugs linked to memory safety, surpassing the count of the second-most prevalent category (logic errors) by a factor of two,
we undertook a more detailed classification of these bugs grounded in their distinct attributes. 
The subcategories we have delineated encompass: Reference Count Misuse (24 cases), Null Pointer Dereference (20 cases), Buffer Overflow (7 cases), Memory Allocation/Releasing (6 cases), Invalid Pointer (3 cases), Stack Overflow (2 cases), and Buffer Permission (1 case).

This subdivision allows us to delve deeper into the various aspects of memory safety issues and develop more specific strategies to address each subcategory. For instance, with Null Pointer Dereference issues, checking whether a pointer is null before dereferencing it is a helpful strategy. For Reference Count Misuse, precise reference count management ensures correct lifecycle management of objects. These specific strategies help with more effectively preventing and fixing memory safety issues, thereby improving the reliability and security of the native Python applications.

These bugs cause erroneous behavior or crashes due to data corruption or invalid memory access. In certain cases, they even pose security threats. 
For instance, an attacker may exploit a buffer overflow in the native code to execute malicious code, or manipulate an invalid pointer to access sensitive data. In other examples, the memory management errors, such as those arising from misuse of reference count or improper memory allocation or release, result in memory leaks or dangling pointers, compromising system performance over time, or leading to unpredictable behaviors.

Figure~\ref{fig:root-memory} depicts a memory-leak bug resulting from improper reference count management in Numpy~\cite{NumPy}. The bug arises when performing a power operation on an array in Python (line p3). Initially, the native function {\tt array\_power} is called, followed by the faulty function invocation {\tt is\_scalar\_with\_conversion} (line c6) along the control flow path underlying this bug.
Carried by the control flow, the Python object {\tt o2}, created from the integer 2 in Python (line p3), eventually reaches the buggy function. At line c8, a new reference to {\tt o2} is established as {\tt value}, which continues to occupy memory even after it is no longer required beyond line c9. This occurs because the reference count has not been decremented. Over time, accumulating such memory leaks consumes monotonically increasing amounts of memory, leading to system performance degradation or crashes (i.e., out of memory). 
To rectify this bug, it is imperative to decrease the reference count of the Python object as soon as it becomes unnecessary in the native code (line c10).

\begin{figure}[tp]
	\vspace{-0pt}
	\includegraphics[width=0.55\linewidth]{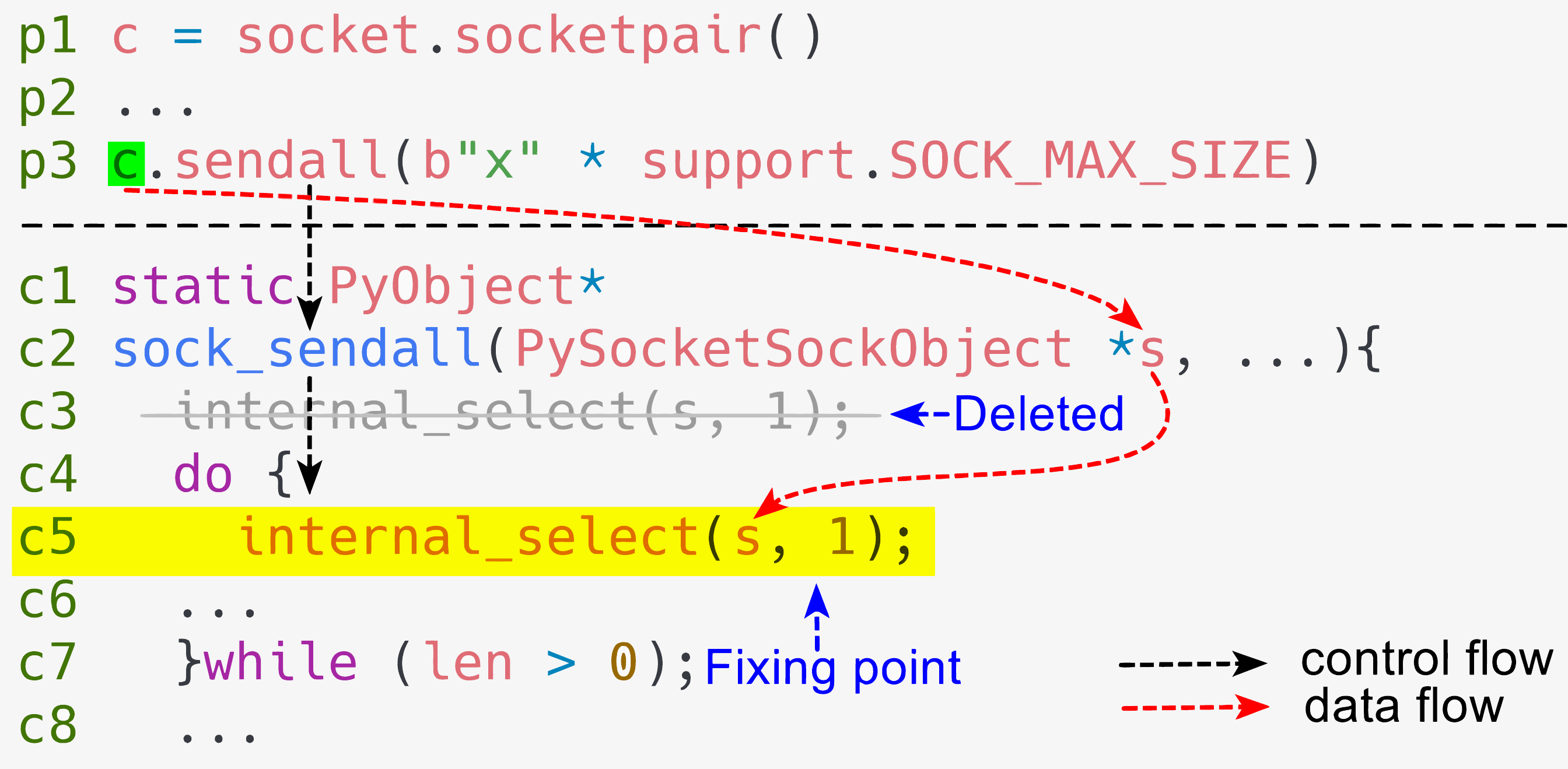}
	\caption{A native bug case rooted in a logic error~\cite{casestudy_logic}.}
	\label{fig:root-logic}
	\vspace{-12pt}
\end{figure}

\underline{{RC2: Logic error (25).}} 
In general, a logic error is a runtime error where a program yields incorrect output or strays from its expected behavior otherwise. 
Such errors often originate from the programmer's misunderstandings or incorrect assumptions while coding. For example, employing an incorrect algorithm occurs when the chosen method to solve a problem is either erroneous or not fully developed. 
In particular, in the context of our studied native code bugs, the bug-causing logic errors are commonly resultant from incorrect understandings and/or assumptions about relevant foreign functions' semantics and the shape of foreign data objects, among other semantic disparities between the two interacting languages (Python and C).

Generally, logic errors do not directly cause a program to fail; however, they result in adverse effects such as incorrect output, data corruption, degraded user experience, and even increased security risks, as demonstrated in our studied bug cases.  
These logic errors have even more significant damages in more severe scenarios, particularly in Python applications dealing with sensitive data. 

As depicted in Figure~\ref{fig:root-logic}, 
a scenario unfolds where a socket timeout is established, 
and an endeavor is made to transmit a quantity of bytes exceeding the immediate transmission capacity. 
In this context, the function prematurely throws an error, failing to adhere to the designated timeout duration. 
To elaborate, if the intention is to transmit 2 bytes, 
yet only one byte can be dispatched instantaneously, 
the function proceeds to send just one byte. 
This anomaly results from positioning the {\tt internal\_select()} call, responsible for monitoring the socket's readiness for transmission, outside the sending loop in its initial implementation.

For addressing this bug, {\tt internal\_select()} invocation is moved into the do-while loop (line c5);
thereby, the function undergoes a transformation, 
compelling it to await the full timeout interval prior to each transmission attempt, thus aligning with the expected behavior.

\underline{{RC3: Data type error (24).}} 
These bugs occur when an operation is performed in a way that is inconsistent with the data type of the value it operates upon. This inconsistency arise from improper data type conversion between Python and C, mismatched data types stemming from assumptions during Python-C integration, or the misuse of data types because of the differences in data type systems across these two languages.

Native code bugs caused by data type errors in our study led to a range of issues, including incorrect program behavior, system crashes, and even the exposure of security vulnerabilities.
Such bugs introduce subtle and elusive defects, making debugging and maintenance endeavors arduous and intricate.
The intricacies of these issues are further amplified, hence the consequences further exacerbated, 
when data objects of types that are incompatible between Python and C have to be involved (e.g., the native code operates on a Python dictionary consisting of elements of heterogeneous data types). 

\begin{figure}[tp]
	\vspace{-0pt}
	\includegraphics[width=0.6\linewidth]{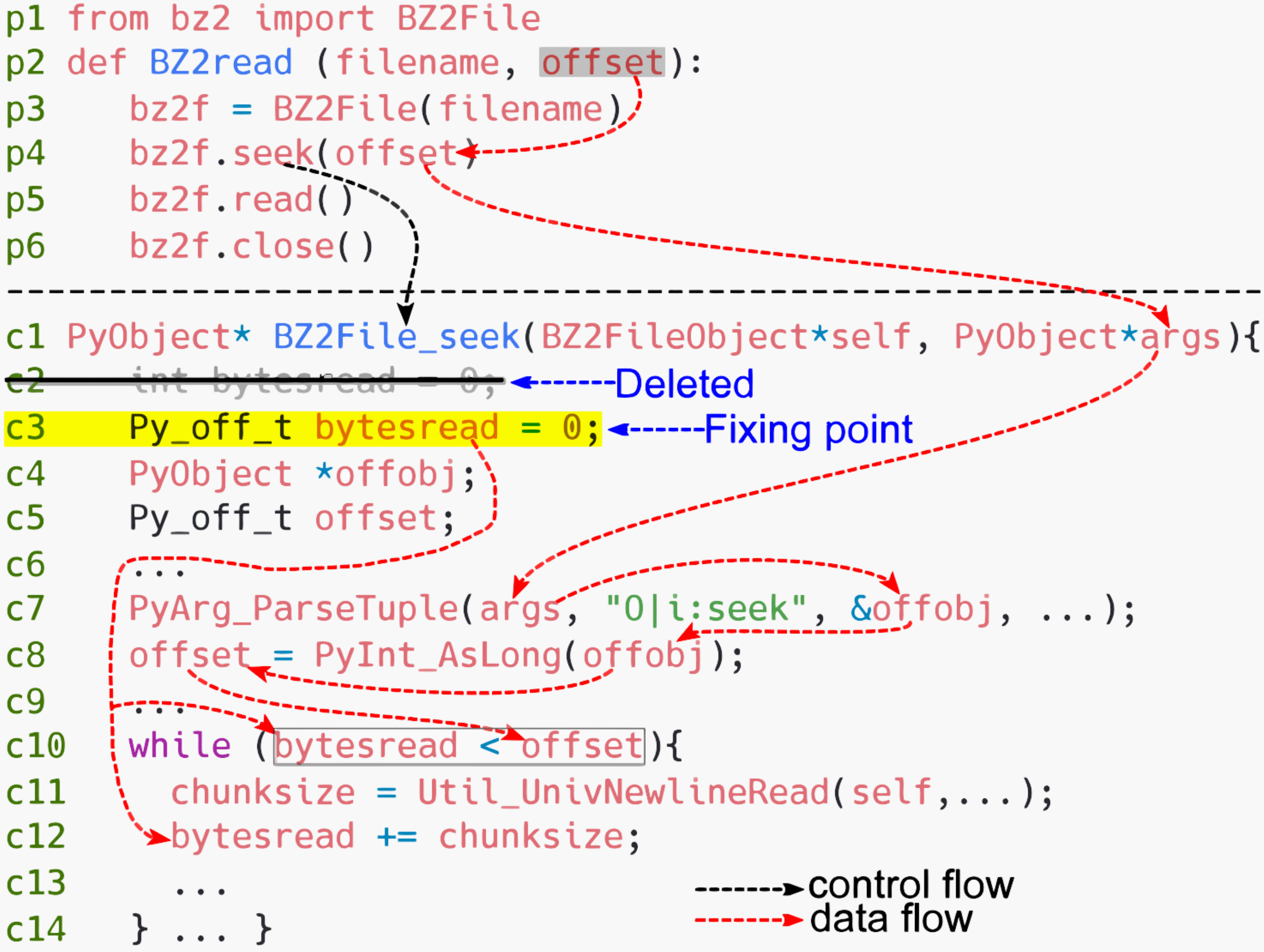}
	\caption{A native bug case rooted in integer overflow~\cite{casestudy_datatype}.}
	\label{fig:root-data-type}
	\vspace{-4pt}
\end{figure}

As depicted in Figure~\ref{fig:root-data-type}, a critical concern emerges when the variable {\tt bytesread} is initially defined as {\tt int} (line c2), while the offset of type {\tt long} is transmitted along the data flow originating from Python, carrying a size surpassing 2GB. In this scenario, a precarious situation arises where an overflow of {\tt bytesread} at line c10 becomes possible, triggering an inexhaustible loop.
This scenario underscores the significance of appropriately managing data types and ensuring compatibility across the language interfaces. An oversight in data type compatibility, as demonstrated in this example, can lead to severe runtime issues, such as infinite loops, compromising the functionality and stability of the system.
To rectify this bug, it is recommended to redefine the variable {\tt bytesread} with the type {\tt Py\_off\_t} (line c3). This adjustment ensures that the two compared variables possess the same data type, and their bit-width aligns with the functional requirements of this Python application.

\begin{figure}[tp]
	\vspace{-4pt}
	\includegraphics[width=0.6\linewidth]{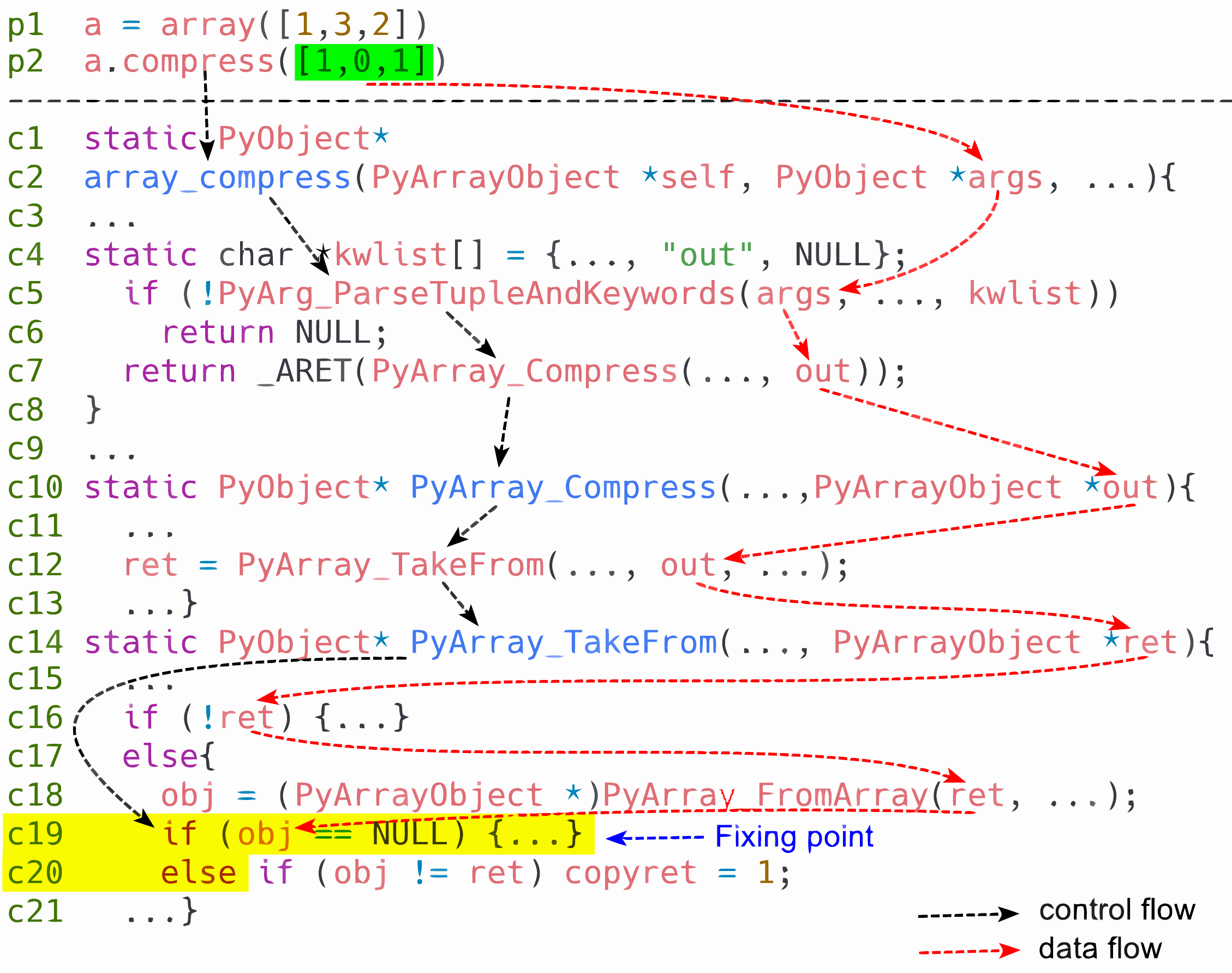}
	\caption{A native bug case rooted in a boundary conditional error~\cite{casestudy_boundary_conditional}.}
	\label{fig:root-boundary}
	\vspace{-8pt}
\end{figure}

\ul{RC4: Boundary conditional error (20).} 
A boundary conditional error, 
such as an off-by-one error or edge case error, arises when a boundary condition within a range is improperly managed. This type of error frequently appears in situations such as array indexing (where an attempt is made to access an array element beyond its boundary), looping constructs (where loop termination conditions are inaccurately defined), and numerical ranges (involving calculations at the extremities of numerical ranges).

These bugs lead to incorrect behavior in programs, causing inaccurate results or unpredictable operation.
In certain cases, these errors cause the program to crash, particularly if they result in attempts to access out of the specific range. Occasionally, these errors cause performance issues by unnecessarily overloading computational resources. 
The potential severity of these outcomes underlines the importance of correctly handling boundary conditions in the native code of Python applications.

In Figure~\ref{fig:root-boundary}, the boundary condition error occurs when the {\tt PyArray\_FromArray} function is called with an array and a data type descriptor that cannot be safely cast to each other. The {\tt PyArray\_FromArray} function is expected to create a new array from an existing array with a specified data type descriptor. However, when the existing array's data type cannot be safely cast to the specified data type descriptor, the function returns NULL (line c18). There was no check to see if {\tt "obj"} was NULL after the call to {\tt PyArray\_FromArray}. This meant that if {\tt PyArray\_FromArray} returned NULL, the code would continue to execute with {\tt "obj"} set to NULL, which leads to a segmentation fault. 
Apparently, the bug is rooted in missing boundary checking, rather than data type errors. 

The fix for the bug was to add a check for this boundary condition. After the call to {\tt PyArray\_FromArray}, the code now checks whether the returned array is NULL. If so, a “ValueError” exception is raised with a message indicating that an array of the proper type fails to be created from the output array, and the function exits with a failure status. This correct handling of the boundary condition prevents the segmentation fault from occurring.

\vspace{0pt}
\find{
The four primary root causes of native code bugs are memory safety violations (29.17\%), logic errors (11.57\%), data type issues (11.11\%), and boundary conditional errors (9.26\%). 
More specifically, the memory safety violations are mainly due to reference count misuse (38.1\%) and null pointer dereference (31.7\%).
%
}

\vspace{-10pt}
\subsection{RQ5: Fixing Strategies of Native Code Bugs}\label{RQ5}
%
%
We further analyzed how the studied native code bugs were fixed based on the 
fixed versions of respective code according to the associated NCBF commits we have confirmed. 
We start by examining the fix locations (\textbf{Fxl}) and then distill the fixing approaches/strategies (\textbf{Fxs}). 

\subsubsection{Fix Location}

Almost all (214) of the 216 native code bugs were fixed within a function. 
In the other two cases, the bug was fixed outside of any 
function---specifically pertaining to changes in function/variable and {\tt struct} declarations. 
This suggests that most of the bug fixes required adjustments within specific functional units of the native code, indicating that the majority of these bugs were contextualized (at the function level) and context-dependent. 
%
In particular, such contexts include the following 7 categories: 

\ul{Fxl1:Foreign function callsite (77).} 
The fixes for these bugs were applied directly at the points where foreign functions were called. The bug-fixings specifically targeted the misuse of the API, arising from discrepancies in data types, conventions, and syntax between the two languages involved.

\underline{Fxl2:If structure (36).} 
These fixes are located within 
conditional structures. Different from the "conditional expression" and "true/false branch" bug-fixing locations (as elaborated below), these fixes involve the whole {\tt if} statement, rather than only a conditional or one of the branches. 
For instance, both the conditional and the {\tt else} body were fixed.

\underline{Fxl3:Conditional expression (27).} 
Fixing these bugs involved modifying a conditional expression. Specifically, the changes addressed the incorrect usage of a variable within a predicate of that expression or resolved an error in the logical operator used.

\underline{Fxl4:True/false branch (19).} 
For these bugs, the bug fixing was limited to one of the branches within an {\tt if} conditional structure. The fix required several code modifications, primarily aimed at rectifying logic errors and correcting incorrect assumptions.

\underline{Fxl5:Other function callsite (19).} 
These bug fixes were applied directly at the call site, although the function being called was not a foreign function. 
Nevertheless, such functions still commonly lie on a control flow path of the native code that is reachable (transitively) from a native function definition and/or 
from a foreign function callsite. 
These fixes mainly addressed incorrect arguments.

\underline{Fxl6:Whole function (12).} 
These fixes were changes to an entire function (i.e., typically across multiple code lines within that function).  

\underline{Fxl7:Assignment (12).} 
The associated bugs were fixed by modifying an assignment statement. The changes specifically focused on correcting the misuse of variable names or addressing issues related to incompatible data types.

While the types of code locations of 
fixes are generally similar to those of bug-introducing locations ($\S$\ref{RQ2}), the distribution of the locations over the bugs is notably different from that over the fixes. 
In particular, while 47 of the bugs were introduced at foreign function callsite, 77 of the bugs were fixed at that kind of locations, meaning many other 
bugs that occurred at other kinds of locations were also fixed by correcting a 
cross-language function call. 
Similarly, fixing conditional expressions enabled patches of not only bugs introduced in those kinds of places but also others. 

\subsubsection{Bug-Fixing Approach}
In examining the various locations of the fixes (i.e., the specific places \textit{where} the corrections were implemented), we also explored the code changes that constituted these fixes (i.e., \textit{how} the fixes were made). We identified eight primary types of these code changes:

\ul{Fxs1:Apply proper memory management or reference counting (76).} 
This strategy addresses memory-related issues, such as memory leaks and improper object lifecycle management. Examples include freeing allocated memory when it is no longer needed to avoid memory leaks. Additionally, this strategy involves incrementing and decrementing reference counts using the appropriate functions (e.g., Py\_INCREF, Py\_DECREF) to ensure proper object lifecycle management when manipulating Python objects in C code.

\ul{Fxs2:Incorporate necessary checks or input validations prior to essential operations (45).} 
This approach includes verifying against NULL pointers, ensuring container (e.g., array) boundaries, and checking against logically invalid values. These precautions are crucial for safeguarding operations like dereferencing pointers, accessing array elements, and performing arithmetic operations (e.g., division).

\underline{Fxs3:Employ appropriate error or exception handling (26).} 
This category of fixes pertains to adding or fixing code logic for handling errors and exceptions. In contrast to the Fxs2 approach, 
which aims to prevent failures in critical operations through the implementation of checks and validations prior to the execution of such operations, this approach focuses on adequately addressing the potential failures \textit{after} the critical operation has occurred. The most prevalent example of this approach within our study was to \textit{change the condition against a foreign function call return value}.

\ul{Fxs4:Utilize suitable data types and casting (17).} 
These fixes' primary changes focused on correcting errors related to data types, notably explicit type conversion (casting). An instance of this is when an integer variable in the native language (C) was used to store floating-point data received from the foreign language (Python), resulting in data corruption issues. The solution was to use the correct data type (i.e., float).

\underline{Fxs5:Correct logical mistakes in conditions or loops (14).} 
This strategy addressed bugs typically stemming from logical errors, as previously identified as a main root cause. A frequent example of this method in our study involved substituting an incorrectly used foreign function with the appropriate one---confirming that a major source of misunderstanding behind the 'logic error' root cause lies in incorrect understandings/assumptions about the semantics of relevant foreign functions (Section~\ref{RQ4}). 

\ul{Fxs6:Improve code structure (11).}
This strategy addresses the maintainability and debugging capabilities of the codebase through better code organization and error reporting. Examples include reorganizing code to improve logical flow and reduce duplication, making it easier to understand and modify, and updating error messages to be more specific and informative, including relevant details such as expected types, or valid range. Additionally, this approach involves removing unused code, variables, and macros to reduce clutter and improve code cleanliness, making the codebase more focused and maintainable.

\ul{Fxs7:Optimize performance (6).}
This strategy addresses the efficiency and resource utilization of the Python-C application. Examples include using more efficient algorithms or data structures to reduce time or space complexity, minimizing unnecessary memory allocations or deallocations, replacing expensive operations with cheaper alternatives when possible, and caching frequently used values to avoid redundant computations.

\ul{Other (21).}
This category encompasses various fix strategies that do not fall under the previously mentioned categories. Examples include implementing workarounds for third-party library bugs and addressing platform-specific issues and compiler differences. These fixes aim to ensure the stability, reliability, and compatibility of the Python-C application across different environments and in the presence of external dependencies.



\vspace{-4pt}
\find{
Almost all (99.1\%) of the native code bugs are fixed within a function and more specifically at foreign function callsites (36.0\%), {\tt If} structure (16.8\%), and conditional expression (12.6\%). 
The specific fixing strategies are diverse in general, among which the most common ones are applying proper memory management or reference counting (35.2\%), incorporating boundary checks or input validations (20.8\%), and employing appropriate error or exception handling (12.0\%)---all typically involving operations on foreign data objects. 
}
\vspace{0pt}
\subsection{RQ6: Challenges of Native Code Bug Detection}\label{RQ6}
On top of the previous analyses of the studied native code bugs (concerning how they are symptomized, manifested, and caused as well as where and how they are fixed), 
we found three main types of detection challenges (\textbf{Cha}) with native code bugs in Python applications. 

{\ul{Cha1: Language-Specific Differences and Interactions (114).}} 
As two heterogeneous programming languages, Python and C each 
has its own unique features, semantics, and idioms. These language-specific differences and how native code interacts with the other 
must be thoroughly understood/considered for detecting the native code bugs. In other words, failure to do so can lead to subtle bugs that are difficult to detect, such as incorrect memory management, or unexpected behavior arising from differences in how these two languages handle certain operations or data structures.  

Accounting for more than half (53\%) of our studied bug cases, this type of challenges represents the dominant barrier to automated native code bug detection techniques, especially those based on static analysis. The analysis/technique would have to reason about the semantics of each language and the interactions between the two languages, including the underlying memory models, which has proven to be challenging~\cite{youn2023declarative,lee2020broadening}. Indeed, semantics disparities between heterogeneous languages are known to be the primary obstacle to multilingual code analysis~\cite{binkley2014orbs,haoran22fsenier}. Now our study shows that this obstacle constitutes the major challenge to detecting most of native code bugs in Python applications, indicating the urgency of overcoming those disparities.

{\ul{Cha2: Foreign Interface Complexity (56).}} 
Native code often interacts with higher-level languages through interfaces or bindings~\cite{li2023multilingual}, which is clearly observed in native Python applications in our study. Bugs in these interfaces can be challenging to detect because they involve both the native and the higher-level language's type safety and error handling, which is also quite true in the native code bug cases we studied. 
For instance, in our cases, improper handling of data types passed between Python and C often led to crashes or unexpected behaviors. The subtleties of these interactions and the potential for mismatches in expectations between the two languages make the detection and diagnosis of native code bugs difficult.

As a result, a static technique to detect native code bugs in Python applications would need to attend to the language interfacing mechanisms and the bindings code, which unfortunately (although justifiably) are diverse~\cite{wen22fsetool,li2023multilingual}. To deal with the underlying issues with type system differences between the languages, static type inference would be a necessary step in the bug detection solution. Yet for a dynamically typed language like Python, precise and complete type inference itself has been a challenging research problem in its own right~\cite{castagna2024polymorphic}. 

{\ul{Cha3: Subtle Logical Errors and Edge Cases (14).}} 
Some native code bugs arise from subtle logical errors or edge cases that may not be immediately apparent during development. These bugs can be difficult to detect because they might not cause obvious crashes or failures but instead lead to incorrect results or unexpected behavior in certain scenarios. Identifying such issues requires a careful analysis of the code logic, considering as many (if not all possible) input combinations and boundary conditions as possible.

The challenge with detecting logic errors is not unique to dealing with native code bugs, but logic-error-induced native code bugs can be even more difficult to detect due to the additional/greater complexity of multilingual code resulting from, again, the semantics disparities between Python and C. Moreover, since such bugs do not come with explicit symptoms and detecting them would need application-specific oracles (e.g., expected program outputs and ground-truth behaviors), even dynamic detection techniques can be greatly deterred~\cite{wen22usenixsecurity,wen23usenixsecurity}---existing dynamic techniques often rely on explicit bug symptoms (e.g., crashes, memory leaks) to recognize the occurrence of logic bugs. 

{\ul{Other (32).}} 
This category encompasses some challenges that have not been previously mentioned, including platform and environment dependencies, as well as error-prone optimizations. These challenges are equally critical for native code bugs, necessitating advanced strategies for precise identification and rectification. 
Yet these challenges are spread over too much to be readily grouped and they do not form any representative category, hence our amalgamation of them all here as a miscellaneous class. 

\vspace{-4pt}
\find{
The primary challenges to detecting native code bugs 
lie in the language-specific differences and interactions (52.8\%), foreign interface complexity (25.9\%), and subtle logical errors and edge cases (6.5\%). 
The first two kinds of challenges are peculiar barriers to static detection techniques, while the last one can also impede dynamic detection techniques. 
}

\section{Discussion}
Based on our findings, 
we provide actionable suggestions to researchers and developers of Python applications on preventing, detecting/testing, and fixing native code bugs.

\subsection{Preventing Native Code Bugs}
Our results ($\S$\ref{RQ1}) revealed that nearly one-third of the studied bugs were symptomized as crashes or abortions. 
While such symptoms are not unique to native code bugs, their strong presence suggests that  \textbf{native code bugs led to critical software failures}. Thus, given the generally greater cost of dealing with defects found later~\cite{mays1990experiences,slaughter1998evaluating}, Python developers should make the best effort possible to prevent these native bugs from happening in the first place. 

\textbf{Use cautions against foreign function calls.} 
In $\S$\ref{RQ2}, we found that the most common location where native code bugs were introduced is foreign function callsite in the native code. This is primarily due to the differences in how each programming language handles data types, memory management, and exceptions, as well as differences in syntax and semantics. Other sensitive locations include callsites of functions that are not cross-language APIs like foreign functions but still involved in cross-language information flow, as well as assignments and return sites that are related to such flows. 
Thus, to avoid introducing native code bugs, Python application developers should pay peculiar attentions to these locations. For instance, a concrete step to be taken is to gain a correct and systematic understanding of a foreign function before invoking it, including its semantics, input/output data types, and return values. 
Now that these foreign functions are provided as part of the Python language implementation (i.e., as Python/C APIs~\cite{pythoncapiweb}) or dedicated binding libraries (e.g., ctypes~\cite{ctypesweb} and SWIG~\cite{swigweb}), they are often well documented. Developers can refer to these official documentation to gain the understanding mentioned. 


\textbf{Need for cross-language API and usage recommendations.}
In our study, we also noticed that language interfacing is complex and diverse: even within the single mechanism of FFI for Python-C software, there are at least four specific instances of such a mechanism (cffi, SWIG, Cython, and ctypes) with each coming with a large and different set of cross-language APIs. 
The complexity and diversity of these scenarios serve as significant obstacles, hindering developers from achieving a complete and accurate understanding of each API. 
Therefore, automated API recommendations could prove very beneficial in this context. Although numerous tools for API recommendations are already available, none of the existing ones address cross-language APIs in Python software development environments. Researchers are thus urged to fill this gap in future studies. 

Considering bugs in native C code in Python applications in particular, the cross-language APIs are mostly foreign functions. 
As noted earlier, as a main portion of cross-language APIs, foreign functions are typically defined and provided by either the language runtime or dedicated third-party (binding) libraries (e.g., ctypes and SWIG). Thus, with respect to a given choice of such FFIs, the foreign functions are definitive and pretty much standardized, facilitating the automated recommendation. 
Recommending correct use of native functions (the other kind of cross-language APIs) would be more challenging because they are user-defined and specific to Python applications. Fortunately, native functions are implemented in native code but expected to be rarely called in native code---in our study, we did not see any invocation of native functions in the native code as relevant to the native code bugs examined.

%
%

\subsection{Testing/Debugging Native Code Bugs}
Our extensive findings on the symptoms, manifestation characteristics, and root causes of native code bugs provide concrete guidance on developing native code bug detection/testing/debugging solutions. 

\textbf{Leverage major symptoms for high-level/first-step detection/testing.}
Our results ($\S$\ref{RQ1}) indicate that native code bugs did often have strong/explicit symptoms such as crashes, error messages, and data corruptions. 
These signals can be leveraged to develop {first-step detection/testing of native code bugs}: e.g., when one of these primary symptoms are observed, it is likely that a native code bug has occurred. 
While this rough indicator alone will not always correctly point to native bugs, 
further examination of the crashes and error messages while looking for evidence of primary root causes of native bugs associated with these symptoms (e.g., memory safety and boundary conditional errors) would help narrow down the scope of inspection---according to the connections between symptoms and root causes we identified. 

\textbf{Exercise key manifestation semantics to trigger native code bugs.}
In $\S$\ref{RQ3}, we revealed that the majority of native code bugs were manifested in 
code semantics related to reference counting, (foreign data) object manipulation, and (cross-language) exception handling. 
Thus, Python application testing techniques targeting native bugs can benefit from focusing on exercising these types of semantics in order to expose/trigger (i.e., manifest) native code bugs. 
For instance, a directed greybox fuzzer can be devised to target respective code entities (e.g., foreign functions for reference management), so as to generate the bug-triggering inputs cost-effectively. Similarly, such bugs can be more efficiently covered by guiding the fuzzer towards exploring exceptional control flow constructs like those handling errors/exceptions (across the host and native languages). 

\textbf{Utilizing common bug-introducing locations for native code bug localization.}
In software debugging, developers would first want to pinpoint where the bug appears (i.e., bug/fault localization). The same applies to native code debugging, for which knowing the most prevalent locations where such bugs are introduced can be immediately helpful. 
For instance, in the general area of software fault localization, deep learning (DL) based techniques have gained increasing attention and successes. 
Leveraging the known, common locations of native code bugs in Python applications, one may develop a DL-based localization technique for such bugs by pre-training and/or fine-tuning (large, pre-trained) code models towards paying particular attention to those locations, so as to increase the model's sensitivity to the most bug-prone entities in the native code of Python applications. 
In fact, a similar idea has proven to be efficacious (albeit for detecting and locating 
cross-language bugs~\cite{haoran24fse}), which benefits from pre-training a Transformer model with an attention mechanism customized in favor of the most common appearing locations of cross-language bugs discovered in an empirical study.

\textbf{Exploit common root causes for causal analysis/diagnosis of native code bugs.}
According to our results in $\S$\ref{RQ4} versus those in $\S$\ref{RQ2}, 
where a native code bug is introduced may not always be where the root cause lies. 
For instance, while a substantial portion of these bugs in our study were introduced at assignment statements related to cross-language behaviors, there may be too many such statements to examine them all. 
Moreover, based on our results in $\S$\ref{RQ5}, only a minor portion of these bugs were fixed by changing assignment statements. Thus, even examining all cross-language information flow related assignments in native code may not effectively lead to locating the real bug. 
Thus, native code fault localization techniques should exploit the major kinds of root causes we discovered towards more effective debugging. 
In particular, given the dominant root cause is memory safety issues, 
the debugger may benefit from memory profiling and sanitization. 

\textbf{Overcome known challenges with detecting native code bugs.}
As summarized in our results for RQ6 ($\S$\ref{RQ6}), detection of native code bugs faces a range of technical challenges, primarily lying in the differences and interactions between varying languages of heterogeneous semantics. 
This is a known barrier in general for multilingual code analysis~\cite{binkley2014orbs}, yet pursuing a language-agnostic approach is both unrealistic and unnecessarily~\cite{haoran22fsenier}. 
On the other hand, as we discussed earlier, purely static analysis based techniques that aim to be precise can be particularly impeded by this primary challenge. Meanwhile, translating one language to another (e.g., the native C code to Python or the other way around) hence converting the multi-language application to a single-language equivalent remains an open research problem, suffering from the lack of a practical level of accuracy and success rate. 
Thus, a more fruitful direction would be investing more in dynamic analysis, with or without help by some static analysis steps---indeed, recent successes of this direction for multi-language software analysis~\cite{wen22usenixsecurity,wen23usenixsecurity}
have supported the promising merits of our recommendation. 

The other dominating challenge lies in the complex interfacing with the foreign language, through Python/C APIs for instance in our study. 
Part of the complexity here is due to how the specific interfacing support (e.g., binding libraries) can be, dealing with which needs to be carefully and comprehensively considered in designing a practical native bug detection technique for multi-language Python software. Further concerning the root causes of the native code bugs that are induced by this challenge suggests that incorporating a type inference module may be beneficial or necessary for achieving the practicality goal here.

\subsection{Fixing Native Code Bugs}
As for bug repair in other software domains, native bug fixing for Python application only comprises of two main steps: locate where the bug-fixing code changes should occur and determine what specific code changes that fix the bug should be made. 
The findings from Section $\S$\ref{RQ5} offer direct guidance on fixing bugs in native code for both steps, through the results on common fix locations and fixing approaches, respectively. 

\textbf{Prioritize foreign function callsite and branching structures for fix localization.}
Our results on common native-bug fix locations suggest that these two types of locations were prevalent in more than half of the bug fixes we analyzed. Therefore, developers may benefit from starting with these areas when seeking efficient solutions to fixing native code bugs, so as to prioritize their debugging efforts. Similarly, automated bug repair methods utilizing deep neural networks, similar to many existing deep-learning-based bug repair tools, could be trained to pay special attention to these locations for more accurate predictions of where repairs should be made. 

Note that among the 216 native bug cases we studied, 21.8\% appeared on foreign function callsites, whereas much more (36\%) of these bugs were fixed by code changes on such callsites. This contrast implies that a non-trivial portion of the bugs that occurred at other types of locations (than foreign function callsites) were also fixed by changing foreign function callsites. At the same time, we indeed found that the bugs that occurred at foreign function callsites were generally fixed by making changes at the same (bug) locations. This further highlights the importance hence the merits of prioritizing foreign function callsites when spotting where to fix native code bugs in Python applications.

On the other hand, in contrast to 29 (13.4\%) of the studied (216) bugs having appeared at assignment statements, only 12 (5.5\%) of the studied bugs were fixed by changing an assignment statements. This implies that a non-trivial portion of the bugs that occurred at assignment statements had to be fixed by changing locations other than assignment statements. 
Thus, for bugs occurring at these statements, developers may want to check 
beyond the bug locations. 

\textbf{Start with the common strategies for native code bug repair.}
In $\S$\ref{RQ5}, we revealed 7 common approaches developers have adopted for fixing native code bugs. 
This result provides clear guidance on what kinds of code changes may be fruitful for writing 
native code bug patches. 
For instance, more than half (121, or 56\%)) of the studied native code bugs were fixed through correcting memory management (e.g., concerning reference counting) or enforcing proper input/boundary checking/validations. 
Knowing the code semantics of these common bug-fixing code changes, along with their dominance over all other fixing strategies, gives developers a practical starting point of thought when figuring out how to fix a native code bug in Python applications. 

It is important to note that when identifying the fixing approaches, we aimed to determine the behaviors of the code changes that fix the root cause, rather than the symptoms, of the respective native code bugs. 
This is essential because it is necessary to fix the root cause of bugs, rather than just refraining their symptoms. 
Thus, following our summarized common fixing strategies, developers will be guided toward fixing root causes of respective bugs. 
For instance, developers may attempt to fix a logic-error-induced native code bug (i.e., fixing the root cause RC2) by identifying a correct cross-language API (more specifically, foreign function) callee to replace the current one at the fault location (i.e., adopting the fixing strategy Fxs5). 
An even more prominent example is that applying proper memory management or reference counting (i.e., the fixing strategy Fxs1) is quite effective at fixing native code bugs arising due to memory safety errors as the root cause.

\section{Threats to Validity}
Our study relies on several tools for data crawling and filtering, which are subject to implementation errors that may have led to incorrect native code bug collection.
One mitigating factor is that these tools have either been widely used by fellow researchers and/or in multiple related studies before. 
To further mitigate this threat, we conducted careful code reviews of 
each tool, especially for the extensions to these tools that we made, and manually validated their correctness against small test samples. 

Also, our study results are subject to manual labeling/analysis errors and human biases. 
To reduce this threat, we followed a rigorous coding process for describing the 
results for each RQ. 
Moreover, we adopted inter-rater and/or negotiated agreement protocols to 
resolve any diverged decisions, which help reduce human biases 
via cross checking. 

Our study results are based on Python-C projects that met our repository mining criteria. 
These projects may not be representative of all existing real-world Python-C software. Accordingly, the bugs we examined do not necessarily represent all of the native code bugs that may occur in Python applications. 
Thus, our findings and conclusions are only best interpreted against the  
projects and native code bugs in them that we have actually studied. 
To mitigate this threat, we considered a sizable number of projects of various application domains while using random sampling with statistically significant sample sizes.
On the other hand, the current scale of our study (i.e., considering 216 bug cases in total) is indeed limited by the lack of more capable mining and analysis tools and the tedious nature of manual analysis. 

Similarly, we cannot claim that our results generalize to non-Python-C software, nor native code bugs in application software other than Python applications. Future work may use our methodology but extend our datasets to address native code bugs in other language combinations.


\section{Related Work}

\noindent
\textbf{Native code analysis}. 
Tan et al.~\cite{tan2021toward} developed a technique for detecting segments within Python code that slow down the execution of a program.
Ma et al.~\cite{ma2023detecting} introduced the PyObject state transition model (PSTM) to monitor and analyze the lifecycle and reference count changes of PyObjects in native code extensions.
Kondoh and Onodera~\cite{kondoh2008finding} utilized a typestate analysis technique along with a coding style checker to inspect Java native code for memory errors across four distinct patterns.
Tan and Croft~\cite{tan2008empirical} applied static analysis and manual inspection to uncover previously undetected bugs, proposing a bug classification taxonomy and systematic solutions to mitigate risks posed by native code.
Li and Tan~\cite{li2009finding,li2011jet} concentrated on identifying improperly managed Java exceptions that originate from native code. 
In comparison to these relevant prior studies, our study is more comprehensive in analyzing native code bugs in multilingual projects (Python-C), addressing bug symptoms, bug locations, bug manifestations, root causes, and fix strategies. 
The previous related studies often address just one or a small subset of these aspects of native bugs, or even just focus on a particular situation in one of the aspects (e.g., reference counting as a particular root cause~\cite{ma2023detecting}).

\vspace{2pt}\noindent
\textbf{Studies on multilingual software}. 
A notable and increasing number of studies have targeted multilingual software, of which many are mainly focused on the diversity~\cite{tomassetti2014empirical} 
and usage of languages~\cite{jones2009software,vasilescu2013babel,steffen2019language,mayer2015empirical} as well as their evolution~\cite{karus2011study,chen2005empirical,li2021understanding}. 
A few others looked further into language companionship~\cite{bissyande2013popularity} and 
linkage/interaction between languages~\cite{mayer2017multi}, as well as 
the effects of multilingual development on the maintenance/comprehension~\cite{kontogiannis2006comprehension,grichi2020impact} or  
quality of resulting software~\cite{kochhar2016large,grichi2020impactjni,abidi2021multi}.
Most of these studies are based on multilingual code, but some of them 
are performed from human-factor perspectives~\cite{abidi2019behind,mayer2017multi}. 
Our study clearly differs from these prior relevant ones in that it focuses on native code bugs in multilingual projects, based on real-world software code and comprehensively examining root causes, fault location patterns, and how to fix the bugs. 

\vspace{2pt}\noindent
\textbf{Studies on code defects}. 
In~\cite{jin2012understanding} the authors examined various aspects (e.g., presence, causes, fixes) of performance bugs in 
GCC, Mozilla, and MySQL. 
Following a similar methodology, 
the study in~\cite{liu2014characterizing} targeted performance bugs too, but concerning those in Android apps. 
Our study also addresses the same aspects, yet with a focus on those in multilingual software (Python applications with native C code in particular)  and a broad scope of bugs instead of performance bugs only. 
Several related studies~\cite{wen22fse,grichi2020impactjni,kochhar2016large} are aimed to discover statistical associations of interest based on potential defect indicators, rather than on actual code bugs. 
And a few others~\cite{berger2019impact,ray2014large,zhang2019study} 
are not really concerned about multilingual code---instead, they mainly addressed single-language software despite looking at a variety of languages. 
In contrast, we study native code defects that have been confirmed as bugs at code level. 


\vspace{2pt}\noindent
\textbf{Studies on issues/bugs in multilingual software}. 
Yang et al.~\cite{yang2023demystifying} manually inspected 586 Stack-Overflow posts related to those issues, while dissecting challenges behind the issues and summarizing current solutions. The study is based on developers' discussions rather than multilingual code. 
Li et al. investigated 
characteristics of bug resolution (e.g., bug open time and reopen rate) in 54 Apache projects~\cite{li2023understandinga} and those of bugs themselves (e.g., code change complexity) in 3 machine-learning frameworks~\cite{li2023understandingb}. 
Like ours, these two studies are based on code artifacts. 
Yet neither the bugs nor the projects were necessarily \textit{multilingual}---they were defined by involving multiple languages without concerning 
language interactions. 
Also, unlike these works, we address code-level root cause, location, manifestation, and fixing strategies of native code bugs in multilingual Python application projects while not being limited to specific software domains.

\section{Conclusion}\label{sec:conclude}
We presented the first comprehensive study of real-world native code bugs 
in the context of (multilingual) Python applications. 
Using our tool dedicated to this study followed by extensive manual inspection and confirmation, we 
collected the first set of 1,039 native code bugs in real-world Python-C projects on GitHub. 
We then conducted in-depth case studies for 216 of these bugs, 
classifying their symptoms, introducing locations, and manifestation characteristics while 
dissecting their root causes and fixing strategies. 
We also identified the key challenges to detecting these native code bugs, primarily from the perspective of developing automated bug-detection techniques. 
Based on our study findings, we provide actionable insights into and practical recommendations on 
how to prevent, detect, test/debug, and fix native code bugs in Python applications.


\balance

\bibliographystyle{ACM-Reference-Format}
\bibliography{paper}


\begin{thebibliography}{82}


\ifx \showCODEN    \undefined \def \showCODEN     #1{\unskip}     \fi
\ifx \showDOI      \undefined \def \showDOI       #1{#1}\fi
\ifx \showISBNx    \undefined \def \showISBNx     #1{\unskip}     \fi
\ifx \showISBNxiii \undefined \def \showISBNxiii  #1{\unskip}     \fi
\ifx \showISSN     \undefined \def \showISSN      #1{\unskip}     \fi
\ifx \showLCCN     \undefined \def \showLCCN      #1{\unskip}     \fi
\ifx \shownote     \undefined \def \shownote      #1{#1}          \fi
\ifx \showarticletitle \undefined \def \showarticletitle #1{#1}   \fi
\ifx \showURL      \undefined \def \showURL       {\relax}        \fi
\providecommand\bibfield[2]{#2}
\providecommand\bibinfo[2]{#2}
\providecommand\natexlab[1]{#1}
\providecommand\showeprint[2][]{arXiv:#2}

\bibitem[git(2020)]%
        {github_api}
 \bibinfo{year}{2020}\natexlab{}.
\newblock \bibinfo{title}{GitHub Developer: provides APIs to retrive or query repositories in GitHub}.
\newblock \bibinfo{howpublished}{\url{https://developer.github.com/v3}}.
\newblock


\bibitem[ten(2024)]%
        {tensorflow}
 \bibinfo{year}{2024}\natexlab{}.
\newblock \bibinfo{booktitle}{\emph{An Open Source Machine Learning Framework for Everyone}}.
\newblock
\urldef\tempurl%
\url{https://github.com/tensorflow/tensorflow}
\showURL{%
\tempurl}
\newblock
\shownote{Last accessed: April 3, 2024}.


\bibitem[Sci(2024)]%
        {SciPy}
 \bibinfo{year}{2024}\natexlab{}.
\newblock \bibinfo{booktitle}{\emph{SciPy library main repository}}.
\newblock
\urldef\tempurl%
\url{https://github.com/scipy/scipy}
\showURL{%
\tempurl}
\newblock
\shownote{Last accessed: April 3, 2024}.


\bibitem[pyt(2024)]%
        {pytorch}
 \bibinfo{year}{2024}\natexlab{}.
\newblock \bibinfo{booktitle}{\emph{Tensors and Dynamic neural networks in Python with strong GPU acceleration}}.
\newblock
\urldef\tempurl%
\url{https://github.com/pytorch/pytorch}
\showURL{%
\tempurl}
\newblock
\shownote{Last accessed: April 3, 2024}.


\bibitem[Num(2024)]%
        {NumPy}
 \bibinfo{year}{2024}\natexlab{}.
\newblock \bibinfo{booktitle}{\emph{the fundamental package for scientific computing with Python}}.
\newblock
\urldef\tempurl%
\url{https://github.com/numpy/numpy}
\showURL{%
\tempurl}
\newblock
\shownote{Last accessed: April 3, 2024}.


\bibitem[Abidi et~al\mbox{.}(2019)]%
        {abidi2019behind}
\bibfield{author}{\bibinfo{person}{Mouna Abidi}, \bibinfo{person}{Manel Grichi}, {and} \bibinfo{person}{Foutse Khomh}.} \bibinfo{year}{2019}\natexlab{}.
\newblock \showarticletitle{Behind the scenes: developers' perception of multi-language practices}. In \bibinfo{booktitle}{\emph{Proceedings of the 29th Annual International Conference on Computer Science and Software Engineering}}. \bibinfo{pages}{72--81}.
\newblock


\bibitem[Abidi et~al\mbox{.}(2021)]%
        {abidi2021multi}
\bibfield{author}{\bibinfo{person}{Mouna Abidi}, \bibinfo{person}{Md~Saidur Rahman}, \bibinfo{person}{Moses Openja}, {and} \bibinfo{person}{Foutse Khomh}.} \bibinfo{year}{2021}\natexlab{}.
\newblock \showarticletitle{Are multi-language design smells fault-prone? an empirical study}.
\newblock \bibinfo{journal}{\emph{ACM Transactions on Software Engineering and Methodology (TOSEM)}} \bibinfo{volume}{30}, \bibinfo{number}{3} (\bibinfo{year}{2021}), \bibinfo{pages}{1--56}.
\newblock


\bibitem[Bae et~al\mbox{.}(2019)]%
        {bae2019towards}
\bibfield{author}{\bibinfo{person}{Sora Bae}, \bibinfo{person}{Sungho Lee}, {and} \bibinfo{person}{Sukyoung Ryu}.} \bibinfo{year}{2019}\natexlab{}.
\newblock \showarticletitle{Towards understanding and reasoning about {Android} interoperations}. In \bibinfo{booktitle}{\emph{2019 IEEE/ACM 41st International Conference on Software Engineering (ICSE)}}. IEEE, \bibinfo{pages}{223--233}.
\newblock


\bibitem[Behnel et~al\mbox{.}(2023)]%
        {cythonsite}
\bibfield{author}{\bibinfo{person}{Stefan Behnel}, \bibinfo{person}{Robert Bradshaw}, \bibinfo{person}{David Woods}, \bibinfo{person}{Matúš Valo}, {and} \bibinfo{person}{Lisandro Dalcín}.} \bibinfo{year}{2023}\natexlab{}.
\newblock \bibinfo{title}{{Cython: C-Extensions for Python}}.
\newblock \bibinfo{howpublished}{\url{https://cython.org/}}.
\newblock


\bibitem[Berger et~al\mbox{.}(2019)]%
        {berger2019impact}
\bibfield{author}{\bibinfo{person}{Emery~D Berger}, \bibinfo{person}{Celeste Hollenbeck}, \bibinfo{person}{Petr Maj}, \bibinfo{person}{Olga Vitek}, {and} \bibinfo{person}{Jan Vitek}.} \bibinfo{year}{2019}\natexlab{}.
\newblock \showarticletitle{On the impact of programming languages on code quality: A reproduction study}.
\newblock \bibinfo{journal}{\emph{ACM Transactions on Programming Languages and Systems (TOPLAS)}} \bibinfo{volume}{41}, \bibinfo{number}{4} (\bibinfo{year}{2019}), \bibinfo{pages}{1--24}.
\newblock


\bibitem[Binkley et~al\mbox{.}(2014)]%
        {binkley2014orbs}
\bibfield{author}{\bibinfo{person}{David Binkley}, \bibinfo{person}{Nicolas Gold}, \bibinfo{person}{Mark Harman}, \bibinfo{person}{Syed Islam}, \bibinfo{person}{Jens Krinke}, {and} \bibinfo{person}{Shin Yoo}.} \bibinfo{year}{2014}\natexlab{}.
\newblock \showarticletitle{ORBS: Language-independent program slicing}. In \bibinfo{booktitle}{\emph{Proceedings of the 22nd ACM SIGSOFT International Symposium on Foundations of Software Engineering}}. \bibinfo{pages}{109--120}.
\newblock


\bibitem[Bissyand{\'e} et~al\mbox{.}(2013)]%
        {bissyande2013popularity}
\bibfield{author}{\bibinfo{person}{Tegawend{\'e}~F Bissyand{\'e}}, \bibinfo{person}{Ferdian Thung}, \bibinfo{person}{David Lo}, \bibinfo{person}{Lingxiao Jiang}, {and} \bibinfo{person}{Laurent R{\'e}veill{\`e}re}.} \bibinfo{year}{2013}\natexlab{}.
\newblock \showarticletitle{Popularity, interoperability, and impact of programming languages in 100,000 open source projects}. In \bibinfo{booktitle}{\emph{2013 IEEE 37th annual computer software and applications conference}}. IEEE, \bibinfo{pages}{303--312}.
\newblock


\bibitem[Bonifro et~al\mbox{.}(2021)]%
        {bonifro2021content}
\bibfield{author}{\bibinfo{person}{Francesca~Del Bonifro}, \bibinfo{person}{Maurizio Gabbrielli}, {and} \bibinfo{person}{Stefano Zacchiroli}.} \bibinfo{year}{2021}\natexlab{}.
\newblock \showarticletitle{Content-Based Textual File Type Detection at Scale}. In \bibinfo{booktitle}{\emph{2021 13th International Conference on Machine Learning and Computing}}. \bibinfo{pages}{485--492}.
\newblock


\bibitem[Castagna et~al\mbox{.}(2024)]%
        {castagna2024polymorphic}
\bibfield{author}{\bibinfo{person}{Giuseppe Castagna}, \bibinfo{person}{Micka{\"e}l Laurent}, {and} \bibinfo{person}{Kim Nguyen}.} \bibinfo{year}{2024}\natexlab{}.
\newblock \showarticletitle{Polymorphic type inference for dynamic languages}.
\newblock \bibinfo{journal}{\emph{Proceedings of the ACM on Programming Languages}} \bibinfo{volume}{8}, \bibinfo{number}{POPL} (\bibinfo{year}{2024}), \bibinfo{pages}{1179--1210}.
\newblock


\bibitem[Chen et~al\mbox{.}(2005)]%
        {chen2005empirical}
\bibfield{author}{\bibinfo{person}{Yaofei Chen}, \bibinfo{person}{Rose Dios}, \bibinfo{person}{Ali Mili}, \bibinfo{person}{Lan Wu}, {and} \bibinfo{person}{Kefei Wang}.} \bibinfo{year}{2005}\natexlab{}.
\newblock \showarticletitle{An empirical study of programming language trends}.
\newblock \bibinfo{journal}{\emph{IEEE software}} \bibinfo{volume}{22}, \bibinfo{number}{3} (\bibinfo{year}{2005}), \bibinfo{pages}{72--79}.
\newblock


\bibitem[Chisnall(2013)]%
        {chisnall2013challenge}
\bibfield{author}{\bibinfo{person}{David Chisnall}.} \bibinfo{year}{2013}\natexlab{}.
\newblock \showarticletitle{The challenge of cross-language interoperability}.
\newblock \bibinfo{journal}{\emph{Commun. ACM}} \bibinfo{volume}{56}, \bibinfo{number}{12} (\bibinfo{year}{2013}), \bibinfo{pages}{50--56}.
\newblock


\bibitem[cournape(2006)]%
        {casestudy_datatype}
\bibfield{author}{\bibinfo{person}{cournape}.} \bibinfo{year}{2006}\natexlab{}.
\newblock \bibinfo{title}{commit of python/cpython}.
\newblock \bibinfo{howpublished}{\url{https://github.com/python/cpython/commit/44b054b84adc2deef45bf3fd2e17532d9a993cd3}}.
\newblock


\bibitem[cournape(2008)]%
        {casestudy_boundary_conditional}
\bibfield{author}{\bibinfo{person}{cournape}.} \bibinfo{year}{2008}\natexlab{}.
\newblock \bibinfo{title}{commit of python/cpython}.
\newblock \bibinfo{howpublished}{\url{https://github.com/numpy/numpy/commit/d8c092816ab6b611692d992cbfaf882a10424793}}.
\newblock


\bibitem[Delorey et~al\mbox{.}(2007)]%
        {delorey2007programming}
\bibfield{author}{\bibinfo{person}{Daniel~P Delorey}, \bibinfo{person}{Charles~D Knutson}, {and} \bibinfo{person}{Christophe Giraud-Carrier}.} \bibinfo{year}{2007}\natexlab{}.
\newblock \showarticletitle{Programming language trends in open source development: An evaluation using data from all production phase sourceforge projects}. In \bibinfo{booktitle}{\emph{Second International Workshop on Public Data about Software Development (WoPDaSD’07)}}.
\newblock


\bibitem[Di~Sipio et~al\mbox{.}(2020)]%
        {di2020multinomial}
\bibfield{author}{\bibinfo{person}{Claudio Di~Sipio}, \bibinfo{person}{Riccardo Rubei}, \bibinfo{person}{Davide Di~Ruscio}, {and} \bibinfo{person}{Phuong~T Nguyen}.} \bibinfo{year}{2020}\natexlab{}.
\newblock \showarticletitle{A Multinomial Na{\"\i}ve Bayesian (MNB) Network to Automatically Recommend Topics for GitHub Repositories}.
\newblock In \bibinfo{booktitle}{\emph{Proceedings of the Evaluation and Assessment in Software Engineering}}. \bibinfo{pages}{71--80}.
\newblock


\bibitem[Grichi et~al\mbox{.}(2020a)]%
        {grichi2020impactjni}
\bibfield{author}{\bibinfo{person}{Manel Grichi}, \bibinfo{person}{Mouna Abidi}, \bibinfo{person}{Fehmi Jaafar}, \bibinfo{person}{Ellis~E Eghan}, {and} \bibinfo{person}{Bram Adams}.} \bibinfo{year}{2020}\natexlab{a}.
\newblock \showarticletitle{On the impact of interlanguage dependencies in multilanguage systems empirical case study on java native interface applications (JNI)}.
\newblock \bibinfo{journal}{\emph{IEEE Transactions on Reliability}} \bibinfo{volume}{70}, \bibinfo{number}{1} (\bibinfo{year}{2020}), \bibinfo{pages}{428--440}.
\newblock


\bibitem[Grichi et~al\mbox{.}(2020b)]%
        {grichi2020impact}
\bibfield{author}{\bibinfo{person}{Manel Grichi}, \bibinfo{person}{Ellis~E Eghan}, {and} \bibinfo{person}{Bram Adams}.} \bibinfo{year}{2020}\natexlab{b}.
\newblock \showarticletitle{On the impact of multi-language development in machine learning frameworks}. In \bibinfo{booktitle}{\emph{2020 IEEE International Conference on Software Maintenance and Evolution (ICSME)}}. IEEE, \bibinfo{pages}{546--556}.
\newblock


\bibitem[gvanrossum(2003)]%
        {casestudy_logic}
\bibfield{author}{\bibinfo{person}{gvanrossum}.} \bibinfo{year}{2003}\natexlab{}.
\newblock \bibinfo{title}{commit of python/cpython}.
\newblock \bibinfo{howpublished}{\url{https://github.com/python/cpython/commit/8f24cdc0d5e30e2f924ed2e8a71400fa87a70983}}.
\newblock


\bibitem[Hu and Zhang(2023)]%
        {hu2023empirical}
\bibfield{author}{\bibinfo{person}{Mingzhe Hu} {and} \bibinfo{person}{Yu Zhang}.} \bibinfo{year}{2023}\natexlab{}.
\newblock \showarticletitle{An empirical study of the Python/C API on evolution and bug patterns}.
\newblock \bibinfo{journal}{\emph{Journal of Software: Evolution and Process}} \bibinfo{volume}{35}, \bibinfo{number}{2} (\bibinfo{year}{2023}), \bibinfo{pages}{e2507}.
\newblock


\bibitem[Hu et~al\mbox{.}(2023)]%
        {hu2023cross}
\bibfield{author}{\bibinfo{person}{Mingzhe Hu}, \bibinfo{person}{Qi Zhao}, \bibinfo{person}{Yu Zhang}, {and} \bibinfo{person}{Yan Xiong}.} \bibinfo{year}{2023}\natexlab{}.
\newblock \showarticletitle{Cross-language call graph construction supporting different host languages}. In \bibinfo{booktitle}{\emph{2023 IEEE International Conference on Software Analysis, Evolution and Reengineering (SANER)}}. IEEE, \bibinfo{pages}{155--166}.
\newblock


\bibitem[Jin et~al\mbox{.}(2012)]%
        {jin2012understanding}
\bibfield{author}{\bibinfo{person}{Guoliang Jin}, \bibinfo{person}{Linhai Song}, \bibinfo{person}{Xiaoming Shi}, \bibinfo{person}{Joel Scherpelz}, {and} \bibinfo{person}{Shan Lu}.} \bibinfo{year}{2012}\natexlab{}.
\newblock \showarticletitle{Understanding and detecting real-world performance bugs}.
\newblock \bibinfo{journal}{\emph{ACM SIGPLAN Notices}} \bibinfo{volume}{47}, \bibinfo{number}{6} (\bibinfo{year}{2012}), \bibinfo{pages}{77--88}.
\newblock


\bibitem[Jones(2009)]%
        {jones2009software}
\bibfield{author}{\bibinfo{person}{Capers Jones}.} \bibinfo{year}{2009}\natexlab{}.
\newblock \bibinfo{booktitle}{\emph{Software engineering best practices}}.
\newblock \bibinfo{publisher}{McGraw-Hill, Inc.}
\newblock


\bibitem[Jones(2010)]%
        {jones2010software}
\bibfield{author}{\bibinfo{person}{Capers Jones}.} \bibinfo{year}{2010}\natexlab{}.
\newblock \bibinfo{booktitle}{\emph{Software engineering best practices: lessons from successful projects in the top companies}}.
\newblock \bibinfo{publisher}{McGraw-Hill Education}.
\newblock


\bibitem[Just et~al\mbox{.}(2014)]%
        {just2014defects4j}
\bibfield{author}{\bibinfo{person}{Ren{\'e} Just}, \bibinfo{person}{Darioush Jalali}, {and} \bibinfo{person}{Michael~D Ernst}.} \bibinfo{year}{2014}\natexlab{}.
\newblock \showarticletitle{Defects4J: A database of existing faults to enable controlled testing studies for Java programs}. In \bibinfo{booktitle}{\emph{Proceedings of the 2014 international symposium on software testing and analysis}}. \bibinfo{pages}{437--440}.
\newblock


\bibitem[Karus and Gall(2011)]%
        {karus2011study}
\bibfield{author}{\bibinfo{person}{Siim Karus} {and} \bibinfo{person}{Harald Gall}.} \bibinfo{year}{2011}\natexlab{}.
\newblock \showarticletitle{A study of language usage evolution in open source software}. In \bibinfo{booktitle}{\emph{Proceedings of the 8th Working Conference on Mining Software Repositories}}. \bibinfo{pages}{13--22}.
\newblock


\bibitem[Kochhar et~al\mbox{.}(2016)]%
        {kochhar2016large}
\bibfield{author}{\bibinfo{person}{Pavneet~Singh Kochhar}, \bibinfo{person}{Dinusha Wijedasa}, {and} \bibinfo{person}{David Lo}.} \bibinfo{year}{2016}\natexlab{}.
\newblock \showarticletitle{A large scale study of multiple programming languages and code quality}. In \bibinfo{booktitle}{\emph{2016 IEEE 23rd International Conference on Software Analysis, Evolution, and Reengineering (SANER)}}, Vol.~\bibinfo{volume}{1}. IEEE, \bibinfo{pages}{563--573}.
\newblock


\bibitem[Kondoh and Onodera(2008)]%
        {kondoh2008finding}
\bibfield{author}{\bibinfo{person}{Goh Kondoh} {and} \bibinfo{person}{Tamiya Onodera}.} \bibinfo{year}{2008}\natexlab{}.
\newblock \showarticletitle{Finding bugs in Java native interface programs}. In \bibinfo{booktitle}{\emph{Proceedings of the 2008 international symposium on Software testing and analysis}}. \bibinfo{pages}{109--118}.
\newblock


\bibitem[Kontogiannis et~al\mbox{.}(2006)]%
        {kontogiannis2006comprehension}
\bibfield{author}{\bibinfo{person}{Kostas Kontogiannis}, \bibinfo{person}{Panos Linos}, {and} \bibinfo{person}{Kenny Wong}.} \bibinfo{year}{2006}\natexlab{}.
\newblock \showarticletitle{Comprehension and maintenance of large-scale multi-language software applications}. In \bibinfo{booktitle}{\emph{2006 22nd IEEE International Conference on Software Maintenance}}. IEEE, \bibinfo{pages}{497--500}.
\newblock


\bibitem[Lee et~al\mbox{.}(2020)]%
        {lee2020broadening}
\bibfield{author}{\bibinfo{person}{Sungho Lee}, \bibinfo{person}{Hyogun Lee}, {and} \bibinfo{person}{Sukyoung Ryu}.} \bibinfo{year}{2020}\natexlab{}.
\newblock \showarticletitle{Broadening horizons of multilingual static analysis: Semantic summary extraction from c code for jni program analysis}. In \bibinfo{booktitle}{\emph{Proceedings of the 35th IEEE/ACM International Conference on Automated Software Engineering}}. \bibinfo{pages}{127--137}.
\newblock


\bibitem[Li and Tan(2009)]%
        {li2009finding}
\bibfield{author}{\bibinfo{person}{Siliang Li} {and} \bibinfo{person}{Gang Tan}.} \bibinfo{year}{2009}\natexlab{}.
\newblock \showarticletitle{Finding bugs in exceptional situations of JNI programs}. In \bibinfo{booktitle}{\emph{Proceedings of the 16th ACM conference on Computer and communications security}}. \bibinfo{pages}{442--452}.
\newblock


\bibitem[Li and Tan(2011)]%
        {li2011jet}
\bibfield{author}{\bibinfo{person}{Siliang Li} {and} \bibinfo{person}{Gang Tan}.} \bibinfo{year}{2011}\natexlab{}.
\newblock \showarticletitle{JET: exception checking in the java native interface}.
\newblock \bibinfo{journal}{\emph{ACM SIGPLAN Notices}} \bibinfo{volume}{46}, \bibinfo{number}{10} (\bibinfo{year}{2011}), \bibinfo{pages}{345--358}.
\newblock


\bibitem[Li and Tan(2014)]%
        {li2014finding}
\bibfield{author}{\bibinfo{person}{Siliang Li} {and} \bibinfo{person}{Gang Tan}.} \bibinfo{year}{2014}\natexlab{}.
\newblock \showarticletitle{Finding reference-counting errors in Python/C programs with affine analysis}. In \bibinfo{booktitle}{\emph{ECOOP 2014--Object-Oriented Programming: 28th European Conference, Uppsala, Sweden, July 28--August 1, 2014. Proceedings 28}}. Springer, \bibinfo{pages}{80--104}.
\newblock


\bibitem[Li et~al\mbox{.}(2022a)]%
        {wen22usenixsecurity}
\bibfield{author}{\bibinfo{person}{Wen Li}, \bibinfo{person}{Ming Jiang}, \bibinfo{person}{Xiapu Luo}, {and} \bibinfo{person}{Haipeng Cai}.} \bibinfo{year}{2022}\natexlab{a}.
\newblock \showarticletitle{{PolyCruise}: A Cross-Language Dynamic Information Flow Analysis}. In \bibinfo{booktitle}{\emph{31st {USENIX} Security Symposium ({USENIX} Security 22)}}.
\newblock


\bibitem[Li et~al\mbox{.}(2022b)]%
        {wen22fse}
\bibfield{author}{\bibinfo{person}{Wen Li}, \bibinfo{person}{Li Li}, {and} \bibinfo{person}{Haipeng Cai}.} \bibinfo{year}{2022}\natexlab{b}.
\newblock \showarticletitle{On the Vulnerability Proneness of Multilingual Code}. In \bibinfo{booktitle}{\emph{ACM Joint Meeting on European Software Engineering Conference and Symposium on the Foundations of Software Engineering (ESEC/FSE)}}.
\newblock


\bibitem[Li et~al\mbox{.}(2022c)]%
        {wen22fsetool}
\bibfield{author}{\bibinfo{person}{Wen Li}, \bibinfo{person}{Li LI}, {and} \bibinfo{person}{Haipeng Cai}.} \bibinfo{year}{2022}\natexlab{c}.
\newblock \showarticletitle{{PolyFax}: A Toolkit for Characterizing Multi-Language Software}. In \bibinfo{booktitle}{\emph{ACM Joint Meeting on European Software Engineering Conference and Symposium on the Foundations of Software Engineering (ESEC/FSE), Tool Demos}}. \bibinfo{pages}{1662--1666}.
\newblock
\urldef\tempurl%
\url{https://doi.org/10.1145/3540250.3558925}
\showDOI{\tempurl}


\bibitem[Li et~al\mbox{.}({[n.\,d.]})]%
        {li2023multilingual}
\bibfield{author}{\bibinfo{person}{Wen Li}, \bibinfo{person}{Austin Marino}, \bibinfo{person}{Haoran Yang}, \bibinfo{person}{Na Meng}, \bibinfo{person}{Li Li}, {and} \bibinfo{person}{Haipeng Cai}.} \bibinfo{year}{[n.\,d.]}\natexlab{}.
\newblock \showarticletitle{How are multilingual systems constructed: Characterizing language use and selection in open-source multilingual software}.
\newblock \bibinfo{journal}{\emph{ACM Transactions on Software Engineering and Methodology}} (\bibinfo{year}{[n.\,d.]}).
\newblock


\bibitem[Li et~al\mbox{.}(2021)]%
        {li2021understanding}
\bibfield{author}{\bibinfo{person}{Wen Li}, \bibinfo{person}{Na Meng}, \bibinfo{person}{Li Li}, {and} \bibinfo{person}{Haipeng Cai}.} \bibinfo{year}{2021}\natexlab{}.
\newblock \showarticletitle{Understanding language selection in multi-language software projects on GitHub}. In \bibinfo{booktitle}{\emph{2021 IEEE/ACM 43rd International Conference on Software Engineering: Companion Proceedings (ICSE-Companion)}}. IEEE, \bibinfo{pages}{256--257}.
\newblock


\bibitem[Li et~al\mbox{.}(2023a)]%
        {wen23usenixsecurity}
\bibfield{author}{\bibinfo{person}{Wen Li}, \bibinfo{person}{Jinyang Ruan}, \bibinfo{person}{Guangbei Yi}, \bibinfo{person}{Long Cheng}, \bibinfo{person}{Xiapu Luo}, {and} \bibinfo{person}{Haipeng Cai}.} \bibinfo{year}{2023}\natexlab{a}.
\newblock \showarticletitle{{PolyFuzz}: Holistic Greybox Fuzzing of Multi-Language Systems}. In \bibinfo{booktitle}{\emph{32nd {USENIX} Security Symposium ({USENIX Security 23})}}.
\newblock
\newblock
\shownote{(artifact evaluated; badges: Available)}.


\bibitem[Li et~al\mbox{.}(2023d)]%
        {li2023pyrtfuzz}
\bibfield{author}{\bibinfo{person}{Wen Li}, \bibinfo{person}{Haoran Yang}, \bibinfo{person}{Xiapu Luo}, \bibinfo{person}{Long Cheng}, {and} \bibinfo{person}{Haipeng Cai}.} \bibinfo{year}{2023}\natexlab{d}.
\newblock \showarticletitle{PyRTFuzz: Detecting Bugs in Python Runtimes via Two-Level Collaborative Fuzzing}. In \bibinfo{booktitle}{\emph{Proceedings of the 2023 ACM SIGSAC Conference on Computer and Communications Security}}. \bibinfo{pages}{1645--1659}.
\newblock


\bibitem[Li et~al\mbox{.}(2023c)]%
        {li2023understandingb}
\bibfield{author}{\bibinfo{person}{Zengyang Li}, \bibinfo{person}{Sicheng Wang}, \bibinfo{person}{Wenshuo Wang}, \bibinfo{person}{Peng Liang}, \bibinfo{person}{Ran Mo}, {and} \bibinfo{person}{Bing Li}.} \bibinfo{year}{2023}\natexlab{c}.
\newblock \showarticletitle{Understanding Bugs in Multi-Language Deep Learning Frameworks}. In \bibinfo{booktitle}{\emph{International Conference on Program Comprehension}}.
\newblock


\bibitem[Li et~al\mbox{.}(2023b)]%
        {li2023understandinga}
\bibfield{author}{\bibinfo{person}{Zengyang Li}, \bibinfo{person}{Wenshuo Wang}, \bibinfo{person}{Sicheng Wang}, \bibinfo{person}{Peng Liang}, {and} \bibinfo{person}{Ran Mo}.} \bibinfo{year}{2023}\natexlab{b}.
\newblock \showarticletitle{Understanding Resolution of Multi-Language Bugs: An Empirical Study on Apache Projects}. In \bibinfo{booktitle}{\emph{ACM/IEEE International Symposium on Empirical Software Engineering and Measurement}}.
\newblock


\bibitem[Liu et~al\mbox{.}(2014)]%
        {liu2014characterizing}
\bibfield{author}{\bibinfo{person}{Yepang Liu}, \bibinfo{person}{Chang Xu}, {and} \bibinfo{person}{Shing-Chi Cheung}.} \bibinfo{year}{2014}\natexlab{}.
\newblock \showarticletitle{Characterizing and detecting performance bugs for smartphone applications}. In \bibinfo{booktitle}{\emph{Proceedings of the 36th international conference on software engineering}}. \bibinfo{pages}{1013--1024}.
\newblock


\bibitem[Ma et~al\mbox{.}(2023)]%
        {ma2023detecting}
\bibfield{author}{\bibinfo{person}{Xutong Ma}, \bibinfo{person}{Jiwei Yan}, \bibinfo{person}{Hao Zhang}, \bibinfo{person}{Jun Yan}, {and} \bibinfo{person}{Jian Zhang}.} \bibinfo{year}{2023}\natexlab{}.
\newblock \showarticletitle{Detecting Memory Errors in Python Native Code by Tracking Object Lifecycle with Reference Count}. In \bibinfo{booktitle}{\emph{2023 38th IEEE/ACM International Conference on Automated Software Engineering (ASE)}}. IEEE, \bibinfo{pages}{1429--1440}.
\newblock


\bibitem[mattip(2021)]%
        {casestudy_reference}
\bibfield{author}{\bibinfo{person}{mattip}.} \bibinfo{year}{2021}\natexlab{}.
\newblock \bibinfo{title}{commit of numpy/numpy}.
\newblock \bibinfo{howpublished}{\url{https://github.com/numpy/numpy/commit/b32b72e3d98d784b98d9c38d4f9905574a60707d}}.
\newblock


\bibitem[Mayer(2017)]%
        {mayer2017taxonomy}
\bibfield{author}{\bibinfo{person}{Philip Mayer}.} \bibinfo{year}{2017}\natexlab{}.
\newblock \showarticletitle{A taxonomy of cross-language linking mechanisms in open source frameworks}.
\newblock \bibinfo{journal}{\emph{Computing}} \bibinfo{volume}{99}, \bibinfo{number}{7} (\bibinfo{year}{2017}), \bibinfo{pages}{701--724}.
\newblock


\bibitem[Mayer and Bauer(2015)]%
        {mayer2015empirical}
\bibfield{author}{\bibinfo{person}{Philip Mayer} {and} \bibinfo{person}{Alexander Bauer}.} \bibinfo{year}{2015}\natexlab{}.
\newblock \showarticletitle{An empirical analysis of the utilization of multiple programming languages in open source projects}. In \bibinfo{booktitle}{\emph{Proceedings of the 19th International Conference on Evaluation and Assessment in Software Engineering}}. \bibinfo{pages}{1--10}.
\newblock


\bibitem[Mayer et~al\mbox{.}(2017)]%
        {mayer2017multi}
\bibfield{author}{\bibinfo{person}{Philip Mayer}, \bibinfo{person}{Michael Kirsch}, {and} \bibinfo{person}{Minh~Anh Le}.} \bibinfo{year}{2017}\natexlab{}.
\newblock \showarticletitle{On multi-language software development, cross-language links and accompanying tools: a survey of professional software developers}.
\newblock \bibinfo{journal}{\emph{Journal of Software Engineering Research and Development}} \bibinfo{volume}{5}, \bibinfo{number}{1} (\bibinfo{year}{2017}), \bibinfo{pages}{1}.
\newblock


\bibitem[Mays et~al\mbox{.}(1990)]%
        {mays1990experiences}
\bibfield{author}{\bibinfo{person}{Robert~G. Mays}, \bibinfo{person}{Carole~L. Jones}, \bibinfo{person}{Gerald~J. Holloway}, {and} \bibinfo{person}{Donald~P. Studinski}.} \bibinfo{year}{1990}\natexlab{}.
\newblock \showarticletitle{Experiences with defect prevention}.
\newblock \bibinfo{journal}{\emph{IBM Systems Journal}} \bibinfo{volume}{29}, \bibinfo{number}{1} (\bibinfo{year}{1990}), \bibinfo{pages}{4--32}.
\newblock


\bibitem[Meyerovich and Rabkin(2013)]%
        {meyerovich2013empirical}
\bibfield{author}{\bibinfo{person}{Leo~A Meyerovich} {and} \bibinfo{person}{Ariel~S Rabkin}.} \bibinfo{year}{2013}\natexlab{}.
\newblock \showarticletitle{Empirical analysis of programming language adoption}. In \bibinfo{booktitle}{\emph{Proceedings of the 2013 ACM SIGPLAN international conference on Object oriented programming systems languages \& applications}}. \bibinfo{pages}{1--18}.
\newblock


\bibitem[Morrissey(1974)]%
        {morrissey1974sources}
\bibfield{author}{\bibinfo{person}{Elizabeth~R Morrissey}.} \bibinfo{year}{1974}\natexlab{}.
\newblock \showarticletitle{Sources of error in the coding of questionnaire data}.
\newblock \bibinfo{journal}{\emph{Sociological methods \& research}} \bibinfo{volume}{3}, \bibinfo{number}{2} (\bibinfo{year}{1974}), \bibinfo{pages}{209--232}.
\newblock


\bibitem[Papamichail et~al\mbox{.}(2016)]%
        {papamichail2016user}
\bibfield{author}{\bibinfo{person}{Michail Papamichail}, \bibinfo{person}{Themistoklis Diamantopoulos}, {and} \bibinfo{person}{Andreas Symeonidis}.} \bibinfo{year}{2016}\natexlab{}.
\newblock \showarticletitle{User-perceived source code quality estimation based on static analysis metrics}. In \bibinfo{booktitle}{\emph{2016 IEEE International Conference on Software Quality, Reliability and Security (QRS)}}. IEEE, \bibinfo{pages}{100--107}.
\newblock


\bibitem[Puzhevich(2020)]%
        {top10language}
\bibfield{author}{\bibinfo{person}{Victoria Puzhevich}.} \bibinfo{year}{2020}\natexlab{}.
\newblock \bibinfo{booktitle}{\emph{Top Programming Languages to Use in 2020}}.
\newblock
\urldef\tempurl%
\url{https://scand.com/company/blog/top-programming-languages-to-use-in-2020/}
\showURL{%
\tempurl}


\bibitem[Python.org(2023)]%
        {ctypesweb}
\bibfield{author}{\bibinfo{person}{Python.org}.} \bibinfo{year}{2023}\natexlab{}.
\newblock \bibinfo{title}{{ctypes — A foreign function library for Python}}.
\newblock \bibinfo{howpublished}{\url{https://docs.python.org/3/library/ctypes.html}}.
\newblock


\bibitem[python.org(2023)]%
        {pythoncapiweb}
\bibfield{author}{\bibinfo{person}{python.org}.} \bibinfo{year}{2023}\natexlab{}.
\newblock \bibinfo{title}{Python/C API Reference Manual}.
\newblock \bibinfo{howpublished}{\url{https://docs.python.org/3/c-api/index.html}}.
\newblock


\bibitem[Ray et~al\mbox{.}(2016)]%
        {ray2016naturalness}
\bibfield{author}{\bibinfo{person}{Baishakhi Ray}, \bibinfo{person}{Vincent Hellendoorn}, \bibinfo{person}{Saheel Godhane}, \bibinfo{person}{Zhaopeng Tu}, \bibinfo{person}{Alberto Bacchelli}, {and} \bibinfo{person}{Premkumar Devanbu}.} \bibinfo{year}{2016}\natexlab{}.
\newblock \showarticletitle{On the" naturalness" of buggy code}. In \bibinfo{booktitle}{\emph{Proceedings of the 38th International Conference on Software Engineering}}. \bibinfo{pages}{428--439}.
\newblock


\bibitem[Ray et~al\mbox{.}(2014)]%
        {ray2014large}
\bibfield{author}{\bibinfo{person}{Baishakhi Ray}, \bibinfo{person}{Daryl Posnett}, \bibinfo{person}{Vladimir Filkov}, {and} \bibinfo{person}{Premkumar Devanbu}.} \bibinfo{year}{2014}\natexlab{}.
\newblock \showarticletitle{A large scale study of programming languages and code quality in github}. In \bibinfo{booktitle}{\emph{Proceedings of the 22nd ACM SIGSOFT International Symposium on Foundations of Software Engineering}}. \bibinfo{pages}{155--165}.
\newblock


\bibitem[Rigo and Fijalkowski(2023)]%
        {cffiweb}
\bibfield{author}{\bibinfo{person}{Armin Rigo} {and} \bibinfo{person}{Maciej Fijalkowski}.} \bibinfo{year}{2023}\natexlab{}.
\newblock \bibinfo{title}{{C Foreign Function Interface for Python (CFFI)}}.
\newblock \bibinfo{howpublished}{\url{https://cffi.readthedocs.io/en/latest/index.html}}.
\newblock


\bibitem[Romano et~al\mbox{.}(2021)]%
        {romano2021g}
\bibfield{author}{\bibinfo{person}{Simone Romano}, \bibinfo{person}{Maria Caulo}, \bibinfo{person}{Matteo Buompastore}, \bibinfo{person}{Leonardo Guerra}, \bibinfo{person}{Anas Mounsif}, \bibinfo{person}{Michele Telesca}, \bibinfo{person}{Maria~Teresa Baldassarre}, {and} \bibinfo{person}{Giuseppe Scanniello}.} \bibinfo{year}{2021}\natexlab{}.
\newblock \showarticletitle{G-Repo: a Tool to Support {MSR} Studies on {GitHub}}. In \bibinfo{booktitle}{\emph{2021 IEEE International Conference on Software Analysis, Evolution and Reengineering (SANER)}}. IEEE, \bibinfo{pages}{551--555}.
\newblock


\bibitem[Slaughter et~al\mbox{.}(1998)]%
        {slaughter1998evaluating}
\bibfield{author}{\bibinfo{person}{Sandra~A Slaughter}, \bibinfo{person}{Donald~E Harter}, {and} \bibinfo{person}{Mayuram~S Krishnan}.} \bibinfo{year}{1998}\natexlab{}.
\newblock \showarticletitle{Evaluating the cost of software quality}.
\newblock \bibinfo{journal}{\emph{Commun. ACM}} \bibinfo{volume}{41}, \bibinfo{number}{8} (\bibinfo{year}{1998}), \bibinfo{pages}{67--73}.
\newblock


\bibitem[Steffen et~al\mbox{.}(2019)]%
        {steffen2019language}
\bibfield{author}{\bibinfo{person}{Bernhard Steffen}, \bibinfo{person}{Frederik Gossen}, \bibinfo{person}{Stefan Naujokat}, {and} \bibinfo{person}{Tiziana Margaria}.} \bibinfo{year}{2019}\natexlab{}.
\newblock \showarticletitle{Language-driven engineering: from general-purpose to purpose-specific languages}.
\newblock In \bibinfo{booktitle}{\emph{Computing and Software Science}}. \bibinfo{publisher}{Springer}, \bibinfo{pages}{311--344}.
\newblock


\bibitem[Sultana et~al\mbox{.}(2016)]%
        {sultana2016understanding}
\bibfield{author}{\bibinfo{person}{Nawrin Sultana}, \bibinfo{person}{Justin Middleton}, \bibinfo{person}{Jeffrey Overbey}, {and} \bibinfo{person}{Munawar Hafiz}.} \bibinfo{year}{2016}\natexlab{}.
\newblock \showarticletitle{Understanding and fixing multiple language interoperability issues: the c/fortran case}. In \bibinfo{booktitle}{\emph{Proceedings of the 38th International Conference on Software Engineering}}. \bibinfo{pages}{772--783}.
\newblock


\bibitem[swig.org(2023)]%
        {swigweb}
\bibfield{author}{\bibinfo{person}{swig.org}.} \bibinfo{year}{2023}\natexlab{}.
\newblock \bibinfo{title}{{SWIG: an interface compiler}}.
\newblock \bibinfo{howpublished}{\url{https://www.swig.org/}}.
\newblock


\bibitem[Tan and Croft(2008)]%
        {tan2008empirical}
\bibfield{author}{\bibinfo{person}{Gang Tan} {and} \bibinfo{person}{Jason Croft}.} \bibinfo{year}{2008}\natexlab{}.
\newblock \showarticletitle{An Empirical Security Study of the Native Code in the JDK}. In \bibinfo{booktitle}{\emph{Usenix Security Symposium}}. \bibinfo{pages}{365--378}.
\newblock


\bibitem[Tan et~al\mbox{.}(2021)]%
        {tan2021toward}
\bibfield{author}{\bibinfo{person}{Jialiang Tan}, \bibinfo{person}{Yu Chen}, \bibinfo{person}{Zhenming Liu}, \bibinfo{person}{Bin Ren}, \bibinfo{person}{Shuaiwen~Leon Song}, \bibinfo{person}{Xipeng Shen}, {and} \bibinfo{person}{Xu Liu}.} \bibinfo{year}{2021}\natexlab{}.
\newblock \showarticletitle{Toward efficient interactions between Python and native libraries}. In \bibinfo{booktitle}{\emph{Proceedings of the 29th ACM Joint Meeting on European Software Engineering Conference and Symposium on the Foundations of Software Engineering}}. \bibinfo{pages}{1117--1128}.
\newblock


\bibitem[Tan et~al\mbox{.}(2017)]%
        {tan2017codeflaws}
\bibfield{author}{\bibinfo{person}{Shin~Hwei Tan}, \bibinfo{person}{Jooyong Yi}, \bibinfo{person}{Sergey Mechtaev}, \bibinfo{person}{Abhik Roychoudhury}, {et~al\mbox{.}}} \bibinfo{year}{2017}\natexlab{}.
\newblock \showarticletitle{Codeflaws: a programming competition benchmark for evaluating automated program repair tools}. In \bibinfo{booktitle}{\emph{2017 IEEE/ACM 39th International Conference on Software Engineering Companion (ICSE-C)}}. IEEE, \bibinfo{pages}{180--182}.
\newblock


\bibitem[Tomassetti and Torchiano(2014)]%
        {tomassetti2014empirical}
\bibfield{author}{\bibinfo{person}{Federico Tomassetti} {and} \bibinfo{person}{Marco Torchiano}.} \bibinfo{year}{2014}\natexlab{}.
\newblock \showarticletitle{An empirical assessment of polyglot-ism in github}. In \bibinfo{booktitle}{\emph{Proceedings of the 18th International Conference on Evaluation and Assessment in Software Engineering}}. \bibinfo{pages}{1--4}.
\newblock


\bibitem[Valverde and Sol{\'e}(2015)]%
        {valverde2015punctuated}
\bibfield{author}{\bibinfo{person}{Sergi Valverde} {and} \bibinfo{person}{Ricard~V Sol{\'e}}.} \bibinfo{year}{2015}\natexlab{}.
\newblock \showarticletitle{Punctuated equilibrium in the large-scale evolution of programming languages}.
\newblock \bibinfo{journal}{\emph{Journal of The Royal Society Interface}} \bibinfo{volume}{12}, \bibinfo{number}{107} (\bibinfo{year}{2015}), \bibinfo{pages}{20150249}.
\newblock


\bibitem[Vasilescu et~al\mbox{.}(2013)]%
        {vasilescu2013babel}
\bibfield{author}{\bibinfo{person}{Bogdan Vasilescu}, \bibinfo{person}{Alexander Serebrenik}, {and} \bibinfo{person}{Mark~GJ van~den Brand}.} \bibinfo{year}{2013}\natexlab{}.
\newblock \showarticletitle{The Babel of software development: Linguistic diversity in Open Source}. In \bibinfo{booktitle}{\emph{International Conference on Social Informatics}}. Springer, \bibinfo{pages}{391--404}.
\newblock


\bibitem[Veeraraghavan(2024)]%
        {top20language}
\bibfield{author}{\bibinfo{person}{Sruthi Veeraraghavan}.} \bibinfo{year}{2024}\natexlab{}.
\newblock \bibinfo{booktitle}{\emph{Top 20 Best Programming Languages To Learn in 2024}}.
\newblock
\urldef\tempurl%
\url{https://www.simplilearn.com/best-programming-languages-start-learning-today-article}
\showURL{%
\tempurl}


\bibitem[Widyasari et~al\mbox{.}(2020)]%
        {widyasari2020bugsinpy}
\bibfield{author}{\bibinfo{person}{Ratnadira Widyasari}, \bibinfo{person}{Sheng~Qin Sim}, \bibinfo{person}{Camellia Lok}, \bibinfo{person}{Haodi Qi}, \bibinfo{person}{Jack Phan}, \bibinfo{person}{Qijin Tay}, \bibinfo{person}{Constance Tan}, \bibinfo{person}{Fiona Wee}, \bibinfo{person}{Jodie~Ethelda Tan}, \bibinfo{person}{Yuheng Yieh}, {et~al\mbox{.}}} \bibinfo{year}{2020}\natexlab{}.
\newblock \showarticletitle{Bugsinpy: a database of existing bugs in python programs to enable controlled testing and debugging studies}. In \bibinfo{booktitle}{\emph{Proceedings of the 28th ACM joint meeting on european software engineering conference and symposium on the foundations of software engineering}}. \bibinfo{pages}{1556--1560}.
\newblock


\bibitem[Yamaguchi et~al\mbox{.}(2014)]%
        {yamaguchi2014modeling}
\bibfield{author}{\bibinfo{person}{Fabian Yamaguchi}, \bibinfo{person}{Nico Golde}, \bibinfo{person}{Daniel Arp}, {and} \bibinfo{person}{Konrad Rieck}.} \bibinfo{year}{2014}\natexlab{}.
\newblock \showarticletitle{Modeling and discovering vulnerabilities with code property graphs}. In \bibinfo{booktitle}{\emph{2014 IEEE Symposium on Security and Privacy}}. IEEE, \bibinfo{pages}{590--604}.
\newblock


\bibitem[Yang et~al\mbox{.}(2022)]%
        {haoran22fsenier}
\bibfield{author}{\bibinfo{person}{Haoran Yang}, \bibinfo{person}{Wen Li}, {and} \bibinfo{person}{Haipeng Cai}.} \bibinfo{year}{2022}\natexlab{}.
\newblock \showarticletitle{Language-Agnostic Dynamic Analysis of Multilingual Code: Promises, Pitfalls, and Prospects}. In \bibinfo{booktitle}{\emph{ACM Joint Meeting on European Software Engineering Conference and Symposium on the Foundations of Software Engineering (ESEC/FSE), Ideas, Visions and Reflections}}. \bibinfo{pages}{1621--1626}.
\newblock
\urldef\tempurl%
\url{https://doi.org/10.1145/3540250.3560880}
\showDOI{\tempurl}


\bibitem[Yang et~al\mbox{.}(2023)]%
        {yang2023demystifying}
\bibfield{author}{\bibinfo{person}{Haoran Yang}, \bibinfo{person}{Weile Lian}, \bibinfo{person}{Shaowei Wang}, {and} \bibinfo{person}{Haipeng Cai}.} \bibinfo{year}{2023}\natexlab{}.
\newblock \showarticletitle{Demystifying Issues, Challenges, and Solutions for Multilingual Software Development}. In \bibinfo{booktitle}{\emph{2023 IEEE/ACM 45th International Conference on Software Engineering (ICSE)}}. IEEE, \bibinfo{pages}{1840--1852}.
\newblock


\bibitem[Yang et~al\mbox{.}(2024a)]%
        {yang2024multi}
\bibfield{author}{\bibinfo{person}{Haoran Yang}, \bibinfo{person}{Yu Nong}, \bibinfo{person}{Shaowei Wang}, {and} \bibinfo{person}{Haipeng Cai}.} \bibinfo{year}{2024}\natexlab{a}.
\newblock \showarticletitle{Multi-Language Software Development: Issues, Challenges, and Solutions}.
\newblock \bibinfo{journal}{\emph{IEEE Transactions on Software Engineering}} (\bibinfo{year}{2024}).
\newblock


\bibitem[Yang et~al\mbox{.}(2024b)]%
        {haoran24fse}
\bibfield{author}{\bibinfo{person}{Haoran Yang}, \bibinfo{person}{Yu Nong}, \bibinfo{person}{Tao Zhang}, \bibinfo{person}{Xiapu Luo}, {and} \bibinfo{person}{Haipeng Cai}.} \bibinfo{year}{2024}\natexlab{b}.
\newblock \showarticletitle{Learning to Detect and Localize Multilingual Bugs}. In \bibinfo{booktitle}{\emph{ACM Joint Meeting on European Software Engineering Conference and Symposium on the Foundations of Software Engineering (ESEC/FSE)}}. \bibinfo{pages}{1--24}.
\newblock
\urldef\tempurl%
\url{https://doi.org/10.1145/3660804}
\showDOI{\tempurl}


\bibitem[Youn et~al\mbox{.}(2023)]%
        {youn2023declarative}
\bibfield{author}{\bibinfo{person}{Dongjun Youn}, \bibinfo{person}{Sungho Lee}, {and} \bibinfo{person}{Sukyoung Ryu}.} \bibinfo{year}{2023}\natexlab{}.
\newblock \showarticletitle{Declarative static analysis for multilingual programs using CodeQL}.
\newblock \bibinfo{journal}{\emph{Software: Practice and Experience}} (\bibinfo{year}{2023}).
\newblock


\bibitem[Zhang et~al\mbox{.}(2019)]%
        {zhang2019study}
\bibfield{author}{\bibinfo{person}{Jie Zhang}, \bibinfo{person}{Feng Li}, \bibinfo{person}{Dan Hao}, \bibinfo{person}{Meng Wang}, \bibinfo{person}{Hao Tang}, \bibinfo{person}{Lu Zhang}, {and} \bibinfo{person}{Mark Harman}.} \bibinfo{year}{2019}\natexlab{}.
\newblock \showarticletitle{A Study of Programming Languages and Their Bug Resolution Characteristics}.
\newblock \bibinfo{journal}{\emph{IEEE Transactions on Software Engineering}} (\bibinfo{year}{2019}).
\newblock


\end{thebibliography}


\end{document}